\documentclass[final,12pt,3p,times,authoryear]{elsarticle}

\usepackage{amssymb}
\usepackage{amsmath}
\usepackage{bm}
\usepackage{booktabs}
\usepackage{float}
\usepackage{placeins}

\begin{document}

\begin{frontmatter}



\title{Elucidating the impact of microstructure on mechanical properties of phase-segregated polyurea: Finite element modeling of molecular dynamics derived microstructures}


\author[inst1]{Steven J. Yang}
\ead{sjy32@cornell.edu}

\affiliation[inst1]{organization={Sibley School of Mechanical and Aerospace Engineering, Cornell University},
            city={Ithaca},
            postcode={14853}, 
            state={NY},
            country={USA}}

\author[inst2]{Stephanie I. Rosenbloom}
\author[inst2]{Brett P. Fors}
\author[inst1]{Meredith N. Silberstein}
\ead{meredith.silberstein@cornell.edu}
\ead[URL]{https://silbersteinlab.com}

\affiliation[inst2]{organization={Department of Chemistry and Chemical Biology, Cornell University},
            city={Ithaca},
            postcode={14853}, 
            state={NY},
            country={USA}}

\begin{abstract}
Phase-segregated polyureas (PU) have received considerable interest due to their use as tough, impact-resistant coatings. Polyureas are favored for these applications due to their mechanical strain rate sensitivity and energy dissipation. Predicting and tailoring the mechanical response of PU remains challenging due to the complex interaction between its elastomeric and glassy phases. To elucidate the role of PU microstructure on its mechanical properties, we developed a finite element modeling framework in which each phase is represented by a volume fraction within a representative volume element (RVE). Critically, we used separate constitutive models to describe the elastomeric and glassy phases. We developed a plasticity-driven breakdown process in which we model the glassy phase disaggregating into a new phase. The overall contribution of each phase at a material point is determined by their respective volume fractions within the RVE. We applied our modeling methods to two compositions of PU with differing elastomeric segment lengths derived from oligoether diamines, Versalink P650 and P1000. Our simulations show that a combination of microstructural differences and elastomeric phase properties accounts for the difference in mechanical response between P650 and P1000. We show our model's ability to predict PU behavior in various loading conditions, including low-rate cyclic loading and monotonic loading over a wide range of strain rates. Our model produces microstructure transformations that mirror those indicated by small-angle X-ray scattering (SAXS) experiments. Fourier transform analysis of our RVEs reveals glassy phase fibrillation due to deformation, a finding consistent with SAXS experiments.
\end{abstract}




\end{frontmatter}



\section{Introduction}

Phase-segregated polyurea elastomers have received considerable attention as a spray-on coating on metal and concrete structures for resisting corrosion, abrasion, and ballistic impact \citep{iqbal_polyurea_2016,shojaei_review_2021,zhang_polyurea_2022}.  Synthesis involves copolymerizing an oligoether diamine (terminated in -N$\text{H}_2$ groups) soft segment with a short diisocyanate hard segment (terminated in -N=C=O groups) forming urea linkages between the segments. The terms soft and hard are used to describe the rubbery or glassy behavior of the segments at ambient temperatures around 293 K. During curing, the urea groups form bidentate hydrogen bonds between hard segments and aggregate into glassy domains. The glassy domains contribute to the mechanical properties by acting as both crosslinks and stiffening fillers. In contrast, the soft rubbery phase consists of a mixture of soft and occluded hard segments. The properties of polyurea can be tailored by changing the lengths of the soft segments. For example, polyurea formulated from Versalink P650 and P1000, two molecular weights of soft segment, have different mechanical properties \citep{guo_experimental_2016,joshi_high_2012,shahi_thermo-mechanical_2021}.

The mechanical properties of polyurea are substantially influenced by its initial microstructure and by microstructural changes during deformation \citep{mott_deformation_2016,runt_phase_2015}. Mechanical testing shows a yield-like stress response and, when unloaded, a large amount of hysteresis with a combination of elastic recovery and permanent set. Upon reloading, the response shows softening, indicating microstructural breakdown. The large amount of hysteresis and softening indicates energy dissipation mechanisms contributing to polyurea’s toughness and impact resistance. The microstructure of polyurea has been extensively studied through atomic force microscopy (AFM) and small-angle X-ray (SAXS) scattering experiments. Tapping mode AFM has shown that the hard phase of polyurea P650 and P1000 assembles into ribbon-like structures \citep{castagna_role_2012,castagna_effect_2013,pangon_influence_2014,amirkhizi_understanding_2022}. Tensile elongation experiments monitored with in-situ AFM reveal that the initially isotropic arrangement of hard phases fragments into rod-shaped domains aligned perpendicular to the loading direction shortly after stretching and then becomes aligned along the loading direction after longer relaxation times \citep{amirkhizi_understanding_2022}. While AFM allows for direct measurement of the structure of polyurea, one drawback is that the images generated by AFM result from cumulative information of a certain depth from the surface
\citep{castagna_role_2012}. Moreover, the acquisition rate for AFM images is on the order of minutes, which restricts measurements to be performed after stress relaxation has occurred \citep{amirkhizi_understanding_2022}.

In contrast, small-angle X-ray scattering experiments provide insight into the microstructural evolution of polyurea under a variety of loading and thermal conditions \citep{balizer_investigation_2011,castagna_effect_2013,choi_microstructure_2012,rinaldi_microstructure_2011,rosenbloom_microstructural_2021}. In situ SAXS during cyclic tensile loading show that after 0.4 true strain, there is a irreversible decrease in hard domain spacing in the loading direction, indicating breakdown \citep{rinaldi_microstructure_2011}. After large deformation, there is an increase in the electron density variance of the scattering peak associated with the hard domain, indicating a decrease in the degree of phase segregation \citep{choi_microstructure_2012}. However, while X-ray scattering provides deep insight into how PU microstructure evolves under various loading conditions, the method averages over millimeter length scales and does not provide molecular force information. 

Several modeling approaches have been utilized to examine the molecular and micro-scale mechanisms underlying the properties of polyurea. The most common methods include fully atomistic molecular dynamics, coarse-grained molecular dynamics, and continuum modeling approaches. Molecular dynamics simulation of polyurea is challenging due to the long-time scales for phase segregation. In addition, the length scale of hard domains (5 – 10 nm interdomain spacing) requires a large RVE, which is difficult to simulate with fully atomistic molecular dynamics. To address these challenges, several authors have simulated the hard and fully intermixed soft phase separately \citep{grujicic_multi-length_2011,heyden_all-atom_2016,manav_molecular_2021}. Another method for generating phase segregation in fully atomistic simulations of PU involves artificially aggregating the hard segments into domains. \citep{grujicic_molecular-level_2011,grujicic_molecular-level_2012}. These artificial methods include algorithms that bring the urea groups closer together, or bend and reposition the chains. Fully atomistic simulations have been successful for examining molecular forces and predicting material properties during high-rate \citep{heyden_all-atom_2016,roy_investigating_2022} and shock loading \citep{manav_molecular_2021,grujicic_molecular-level_2011,grujicic_molecular-level_2012}. Simulations reveal key molecular contributions of the hard phase. During shock loading, there is a substantial breaking of hydrogen bonds, which contributes to energy dissipation and densification \citep{manav_molecular_2021,grujicic_molecular-level_2012}. The coarse-grained molecular dynamics (CGMD) approach allows for simulating longer times and larger length scales. The coarse-grained approach involves representing monomers or groups of connected atoms as individual beads, considerably reducing computational expense. Simulated equilibration of coarse-grained polyurea can generate phase-segregated morphologies \citep{agrawal_simultaneous_2014,zheng_molecular_2022}. In addition, coarse-grained simulations have successfully predicted PU mechanical properties for intermediate and low-rate loading. \citet{agrawal_prediction_2016} was able to obtain good agreement for storage and loss modulus between coarse-grained PU and ultrasonic experiments for frequencies between $10^3$ rad/s and $10^6$ rad/s. \citet{zheng_molecular_2022} achieved good agreement between coarse-grained polyurea and monotonic tensile experiments at a strain rate of 0.005 $\text{s}^{-1}$. The study analyzed the microscale contributions of stress and found that initially, the hard domains were the primary contributor to total stress, while at larger strains, the soft phase became dominant. Moreover, the study noted that the hard phase is co-continuous after equilibration but then fractures into discrete fragments during loading.

Continuum modeling approaches can enhance conceptual understanding of mechanical behavior. Most polyurea continuum models can be categorized into two types: viscoelastic models and micromechanical models inspired by microstructure. Viscoelastic models have been successful in predicting stress-strain for monotonic tension and compression loading across a wide range of strain rates from quasistatic to high-rate \citep{amirkhizi_experimentally-based_2006,bai_hyper-viscoelastic_2016,chen_tensile_2022,gamonpilas_non-linear_2012,li_hyper-viscoelastic_2009}. Certain micromechanical polyurea models represent the hard and soft phases as different model elements \citep{cho_constitutive_2013,shim_rate_2011}. In these models, linear elastic, hyperelastic, and viscoplastic constitutive relations are specified to represent the mechanics of each phase. These models successfully capture stress-strain for both monotonic and cyclic loading. Micromechanical models homogenize the properties of heterogeneous materials. Thus, modeling polyurea requires specifying the kinematic relationship between the hard and soft phases. The studies above specify that equal deformation is applied to the mechanisms representing the two phases. Certain micromechanical models consider the heterogeneous strain distributions of phase-segregated polymers. \citet{qi_stressstrain_2005} developed a micromechanical model for phase-segregated polyurethane where the chain stretch of the soft phase is adjusted by a strain amplification factor dependent on the hard phase volume fraction. Yet, while these continuum models effectively capture PU mechanical responses over a wide range of loading conditions, they do not explicitly describe how strain distributions throughout the microstructure influence overall mechanical behavior.

For heterogeneous materials, explicit representation of structure within a finite element framework is effective for understanding how each component influences bulk properties. These simulations are typically performed with a representative volume element (RVE) for which sufficient microstructural features are included to yield results representative of the bulk material \citep{david_muzel_application_2020,uthale_polymeric_2021}. The finite element RVE method has been used to study a variety of heterogeneous materials, including polymer nano-composites \citep{mortazavi_modeling_2013,sadati_experimental_2022}, particle-filled mycelium-based composites \citep{islam_mechanical_2018}, and phase-segregated copolymers \citep{grujicic_multi-length_2011,tzianetopoulou_micromechanics_2003,zeleniakiene_simulation_2018,zhang_impact_2020}. Specific to polyurea, \citet{grujicic_multi-length_2011} performed a Mesodyn dynamic mesoscale simulation of PU phase segregation and then converted the resulting morphology into a finite element mesh. The elements within the mesh were assigned properties corresponding to either the soft phase (modeled as hyperelastic) or the hard phase (modeled as elastic-plastic). Material parameters for each phase were obtained through separate, fully atomistic MD simulations. When comparing the stress-strain curve of the finite element RVE with experimental loading and unloading cycles for P1000, the authors successfully captured the loading behavior of PU, including its yield-like characteristics and the loss of stiffness at larger strains caused by non-linear material and geometric effects. \citet{zhang_impact_2020} used tapping mode AFM images as geometric inputs for mesoscale finite element simulation of polyurea. In addition to the hard and soft phase geometry extracted from the AFM images, the authors artificially generated interfacial regions of various thicknesses. The elements were assigned properties of the soft phase, the hard phase, and the interface. Results from the study elucidate how the properties of the individual phases, the presence of an interfacial region, and the relative volume fractions of the phases, impact overall viscoelastic behavior.

The key role of microstructure in PU mechanical properties is a central assertion in the literature and has strong implications for the design of tough elastomers. In this study, we explored how the microstructure of polyurea affects its mechanical properties. We propose that the influence of the soft segment chain length on PU's mechanical properties can be understood by an explicit representation of domain microstructure and its subsequent evolution. Specifically, we used coarse-grained molecular dynamics simulations to generate microstructure for two well-studied PU formulations of different soft segment lengths (P650 and P1000). We then applied the resulting microstructure to finite element simulations where each phase is described with separate constitutive models. Critically, the hard phase constitutive model contains plasticity-driven breakdown into soft phase-like properties. We examined the model behavior over various loading histories and compared the results with both macroscale and microstructural experimental data. This manuscript has the following sections: (\ref{sec:methods_rve_construction}) Modeling methods for RVE construction and simulation. Experimental methods for (\ref{sec:PU_synthesis}) PU synthesis, (\ref{sec:PU_mechanical_characterization}) mechanical testing, and (\ref{sec:PU_SAXS}) in situ small-angle X-ray scattering during deformation. (\ref{sec:constitutive_model}) Constitutive theory for each phase. (\ref{sec:results_md}) Molecular dynamics and RVE generation results. (\ref{sec:results_parameter_id}) Constitutive model parameter identification. (\ref{sec:results_fe_exp}) Results and discussion exploring the model and comparison with experiments. (\ref{sec:conclusion}) Conclusion.

\section{Methods}
\subsection{Modeling methods for RVE construction and simulation}
\label{sec:methods_rve_construction}

\begin{figure}[htb!]
	\centering
		\includegraphics[scale=0.75]{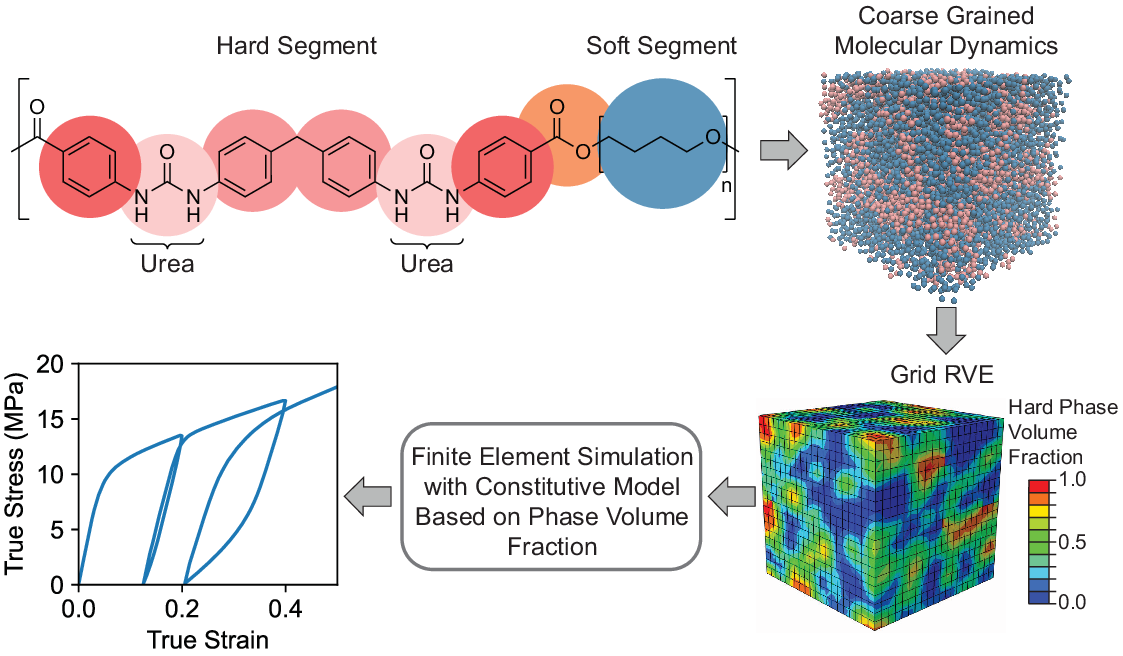}
	\caption{A schematic outlining our simulation process. We represented the chemical structure of polyurea (PU) as coarse-grained beads and simulated phase segregation using Coarse-Grained Molecular Dynamics (CGMD). The obtained data was then converted into a volume fraction grid, acting as the Representative Volume Element (RVE) for subsequent finite element simulations. By applying constitutive equations to the hard, soft, and broken-down phases using the RVE's volume fraction data, we simulated the impact of microstructure on the behavior of PU using finite element analysis.}
	\label{FIG:Methods_Flow_1}
\end{figure}

A schematic of our procedure for modeling polyurea microstructure as a finite element representative volume element (RVE) is presented in Figure \ref{FIG:Methods_Flow_1}. Specifically, we represented the glassy and elastomeric phases as regions discretized by a finite element mesh. This was achieved by assigning each material point a volume fraction of glassy and elastomeric phases. Our constitutive model describing the phases is presented separately from this methods section in Section \ref{sec:constitutive_model}: Constitutive Model. In Section \ref{sec:methods_cgmd}, we describe creating the microstructure for our RVEs using coarse-grained molecular dynamics (CGMD). We generated coarse-grained representations for polyurea P650 and P1000 and performed a simulated annealing process. In Section \ref{sec:methods_cgmd_2_fe_rve}, we describe using the resulting microstructure for CGMD to generate a finite element RVE. A volumetric density map of the coarse-grained beads was created to assign the hard phase volume fractions of the finite element material points. Section \ref{sec:methods_fe} outlines the methods used in our finite element simulations of a 10x10x10 $\text{nm}^{\text{3}}$ RVE to evaluate the stress-strain response of our model subjected to various loading conditions. In Section \ref{sec:fe4ft} we describe finite element simulations of a larger 60x60x60 $\text{nm}^{\text{3}}$ RVE. We performed Fourier transforms on the larger RVE to track the evolution of the simulated microstructure in order to compare our simulations with results from small-angle X-ray scattering experiments.

\subsubsection{Coarse-grained molecular dynamics simulations}
\label{sec:methods_cgmd}

We employed the coarse-grained potentials developed by \citet{agrawal_prediction_2016} to generate phase-segregated structures of simulated P650 and P1000. In this approach, polyurea chains are represented by five different bead types: one bead type for the (-C4H8O-) monomer in the soft segment and four different bead types, representing half of the symmetric aromatic hard segment, for a total of eight beads per hard segment. The bond stretch and angle energies between beads are modeled using harmonic potentials, while non-bonded interaction energies are modeled using tabular data. We determined the number of beads in each soft segment by characterizing the molecular weight of P650 and P1000 oligomers using THF gel permeation chromatography. The soft segment molar mass of P650 and P1000 is 952.7 g/mol and 1659.7 g/mol, respectively (Supplementary Materials Section S1). As each soft segment bead has a molar mass of 72.10 g/mol, our simulated P650 and P1000 contain 13 and 23 beads per soft segment, respectively. \citet{agrawal_prediction_2016} found that modeling bulk polyurea as monodisperse with seven alternating hard-soft blocks accurately captures its thermomechanical properties. Accordingly, our simulated P650 chain consists of $(\text{S}_{6} \text{H}_{2} \text{S}_{7})_{7}$, and our simulated P1000 chain consists of $(\text{S}_{11} \text{H}_{2} \text{S}_{12})_{7}$, where S and H represent the soft and hard repeat units, respectively.

For computational efficiency, we employed two different system sizes. This was because the resulting microstructure served as an input for finite element simulations, which require significantly more time as system size increases. For the simulations of various loading histories, we chose to use a periodic system size of approximately 10 nm in each dimension, which is able to include several hard domains within the volume. The 10 nm system size corresponds to 57 chains for P650 and 38 chains for P1000. In order to analyze the evolution of the hard phase structure, we conducted spatial frequency analyses on RVEs measuring 60 nm in each dimension. For computational expediency, we conducted our coarse-grained molecular dynamics simulation on a system with dimensions of 40 nm. This translates to 1532 chains for P650 and 1038 chains for P1000. Subsequently, we expanded our RVE size to 60 nm for our finite element simulations.

We performed the following simulation steps using the Large-scale Atomic/Molecular Massively Parallel Simulator (LAMMPS) program on Stampede2 (ACCESS) \citep{LAMMPS}. First, we assembled random walk chains at a low density of 0.25 $\text{g}/\text{c}\text{m}^{3}$ in a periodic simulation cell. To relax initial high-energy configurations, we performed a run using a Langevin thermostat at 500 K using soft pairwise potentials. Next, using the harmonic and non-bonded potentials developed by \citet{agrawal_prediction_2016}, we performed an NVT run (Nose-Hoover thermostat) at 300 K where we compressed the simulation cell to a density of ~1 $\text{g}/ \text{c} \text{m}^{3}$ for 0.01 ns. Last, to generate a phase-segregated structure, we performed an NPT run (Nose-Hoover thermostat and barostat) with a temperature of 300 K and pressure of 1 atm for 100 ns. After 100 ns, the density increased to $\sim$1.08 and $\sim$1.02 g/mol for P650 and P1000, respectively.

\subsubsection{Generating RVEs for finite element simulations}
\label{sec:methods_cgmd_2_fe_rve}

We created RVEs for finite element simulations by generating volumetric maps based on the MD equilibrated configurations. The volumetric map provides a continuous representation of the bead number density, allowing for flexible discretization of the finite element RVE. The volumetric map was generated by substituting each bead in the MD results with a Gaussian distribution of approximately the bead radius (3 {\AA} standard deviation). To make the RVE periodic, we incorporated one periodic image beyond every face and corner of the simulation cell. The sum of the Gaussian distributions gives a number density for the soft and hard beads, which was used to generate a volume fraction of the hard phase at each material point for the finite element RVE. 

\subsubsection{Finite element simulation for mechanical behavior}
\label{sec:methods_fe}
Finite element (FE) simulations were performed using Abaqus/Explicit (Dassault Systemes) with a user-defined material subroutine (VUMAT) to implement our constitutive model. For the mesh, we used 3D reduced integration hexahedral elements (C3D8R),  specifying 0.5 nm side lengths for each element. The finite element RVE is a cube with a 10 nm side length for simulated low-rate cyclic and high-rate loading. Periodic boundary conditions were employed using the dummy node method \citep{garoz_consistent_2019}. As Abaqus/Explicit requires all nodes to be linked to the body, we designated three corner nodes of the model as dummy nodes. Any other boundary constraints associated with these nodes were eliminated. Due to the large difference in stiffness between hard and soft phases and the relatively coarse mesh, we used the arbitrary Lagrangian-Eulerian (ALE) remeshing feature in Abaqus to prevent excessive distortion of the elements. We used the volume smooth method with one remeshing sweep every 10 steps. Uniform mass scaling was used to reduce computational time for the explicit simulations; we verified that the amount of kinetic energy was less than 5\% of the total energy after the initial few loading steps. At the start of each simulation, the hard phase volume fraction was stored as a state variable at each integration point, creating a phase-segregated morphology. Uniaxial tension and compression are modeled by displacing one dummy node along an axis while constraining the other two to a plane with the freedom to contract or expand perpendicularly. Low-rate cyclic simulations involved tension loading to 0.2 true strain, unloading to zero force, loading to 0.4 true strain, unloading, and loading to 0.55 true strain. High-rate simulations involved monotonic tension or compression to 0.5 true strain.

\subsubsection{Finite element simulation for analysis of hard phase structure}
\label{sec:fe4ft}

We conducted separate finite element simulations with a larger 60 nm RVE for microstructural analysis. The larger RVEs improved the frequency resolution of our discrete Fourier transform analyses. We used the same Abaqus/Explicit with VUMAT setup as discussed above (Section \ref{sec:methods_fe}), along with the same element type and ALE remeshing feature. Because of computational limitations, we applied homogeneous deformation boundary conditions instead of the periodic boundary conditions used for the smaller RVEs. This condition ensures the RVE faces remain planar with displacement boundary conditions on three faces and nodal kinematic constraints of the three free faces. The larger RVE was then subjected to monotonic tension loading up to a strain of 0.55.

We used a 3D discrete Fourier transform to examine how the hard domain morphology of the RVE simulation changes during deformation. This analysis was done using the fast Fourier transform (FFT) tool in SciPy \citep{2020SciPy-NMeth}. Since the FFT approach requires evenly spaced data points, and the RVE deforms into a non-regular shape, we interpolated the RVE data onto a regular 3D grid. We used a nearest-neighbor interpolation for its computational efficiency. The Fourier transform, performed on a spatial domain, generates a spatial frequency image in Fourier space. To study the frequency of hard phase clusters and their domain spacing, we examined a slice from the center of the Fourier space. In this slice, the vertical axis indicates the domain frequency along the loading direction, and the horizontal axis is across the loading direction. We reduced the pattern to 1D through full azimuthal integration and integration of 30-degree slices aligned with loading and transverse directions. We then manually fitted the 1D scattering peak to either a normal or log-normal distribution to determine domain spacing for each 1D pattern. 

\subsection{Experimental methods for PU synthesis, mechanical testing, and in situ small-angle X-ray scattering during deformation}
\label{sec:experimental_methods}
While several publications report low and high-rate monotonic loading experiments for P650 and P1000, we found a gap in published data for low-rate cyclic loading for P650 and P1000. In addition, despite the abundant literature data characterizing P1000 through small-angle X-ray scattering (SAXS), we identified a notable absence of in situ SAXS measurements for P650 during tensile deformation. Thus, we synthesized PU P650 and P1000 and performed the mechanical tests and X-ray characterization described in this section.

\subsubsection{PU synthesis and specimen preparation}
\label{sec:PU_synthesis}
We synthesized the two compositions, P650 and P1000, from a  polycarbodiimide-modified diphenylmethane diisocyanate (Isonate 143L, Dow) and an oligoether diamine (Versalink P650 and P1000, Evonik). Synthesis of polyurea is typically performed with a stoichiometric excess of diisocyanate. Therefore, we mixed the components at a weight ratio of 1:2.82 for P650 and 1:3.95 for P1000 to achieve a 105\% stoichiometric ratio. We added ~2-3 mL of tetrahydrofuran per 5 g of Versalink to thin the mixture for drop casting. The mixture was heated on a hotplate at 60 $^\circ$C and mixed with a magnetic stirrer for $\sim$3 minutes. The mixture was drop cast in a flat bottom Teflon dish and allowed to solidify over 24 h at room temperature. The resulting polymer was then removed from the dish and annealed in an oven at 80 $^\circ$C under vacuum for 48 hrs. The cured polymers were cut into ASTM D638-V dog-bone shapes with a custom die cutter. A speckle pattern was applied to the dog-bone specimens using black spray paint for digital image correlation strain measurements (DIC).

\subsubsection{Mechanical testing}
\label{sec:PU_mechanical_characterization}
We performed cyclic tensile loading on a universal testing machine (Zwick/Roell Z010) with a 500 N load cell. Digital image correlation strain measurements were performed using a CCD camera (Point Grey) with a 150 mm macro lens (Nikon) and software by Correlated Solutions. We found the effective gauge length for the dog-bone geometry to be 19.54 mm. The specimens were stretched at a rate of 1.95 mm/s, 0.195 mm/s, 0.0195 mm/s for the 0.1 $\text{s}^{\text{-1}}$, 0.01 $\text{s}^{\text{-1}}$, and 0.001 $\text{s}^{\text{-1}}$ strain rate experiments. For the loading cycles, we stretched the specimens to true strain values based on the effective gauge length of the dog-bone geometry. The cyclic loading experiments involved the following steps: we stretched the specimen to 0.2 true strain, unloaded to zero force, stretched the specimen to 0.4 true strain, unloaded to zero force, and then stretched the specimen to 0.55 true strain. We averaged the strains in a rectangular region of the gauge area to obtain strain measurements for the loading and transverse directions. To calculate true stress, we assumed that the out-of-plane strain equals the measured transverse strain to calculate the change in the cross-sectional area of the gauge section during deformation.

\subsubsection{In situ small-angle X-ray scattering during deformation}
\label{sec:PU_SAXS}
Experiments were conducted at the Functional Materials Beamline (FMB) at the Cornell High Energy Synchrotron Source (CHESS) to observe in situ small-angle X-ray scattering during tensile loading. Experiments were conducted with X-ray energy of 15.89 keV which was selected using a diamond (220) side bounce monochromator. The X-ray beam size was approximately 0.3x0.3 $\text{mm}^{\text{2}}$, and it was adjusted using slits positioned approximately 1 meter away from the tensile stage. A Pilatus 300K detector was placed approximately 2 meters downstream from the tensile stage with a helium flight path. One second exposure times were used throughout the experiment. The specimens were stretched at a strain rate of 0.01 $\text{s}^{\text{-1}}$ with a portable tensile stage (Linkham Scientific). We used the BioXTAS RAW software to perform background subtraction and azimuthal integration to generate 1D scattering profiles \citep{hopkins_bioxtas_2017}. We performed full azimuthal integration, as well as integration over 30-degree slices in both the loading and transverse directions. 

\section{Constitutive model}
\label{sec:constitutive_model}
\begin{figure}[htb!]
	\centering
		\includegraphics[scale=1.0]{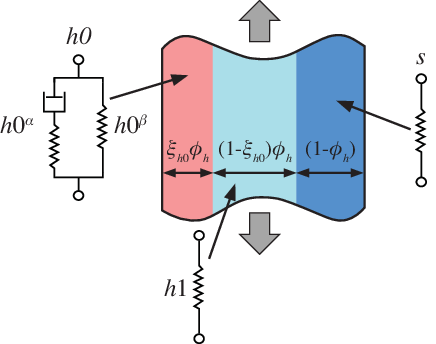}
	\caption{Rheological schematic of model mechanisms at each material point. The red region represents the initial hard phase (mechanism $h0$), the light blue region represents the broken-down hard phase (mechanism $h1$), and the dark blue region represents the soft phase (mechanism $s$). The proportion that mechanisms $h0$, $h1$, and $s$ contribute to the response are controlled by state variables $\xi_{h0}$ and $\phi_h$.}
	\label{FIG:Model_1}
\end{figure}

We model the response of each material point as a weighted combination of mechanisms representing the soft and hard phases. A schematic of the rheological elements of each mechanism and how the mechanisms combine to form the total response is shown in Figure \ref{FIG:Model_1}.
Initially, the mechanical response is from mechanism $h0$, which represents the hard phase before breakdown, and mechanism $s$, which represents the soft phase. As the material is deformed, we model the broken-down hard phase as mechanism $h1$. Mechanism $h0$ is an elastic-viscoplastic glassy polymer model. A linear-elastic element in series with a nonlinear dashpot (mechanism $h0^{\alpha}$) captures the stiffness and yielding of the intermolecular interactions. A parallel neo-Hookean element (mechanism $h0^{\beta}$) captures hardening after yield attributed to resistance due to network reorientation. In parallel, mechanism $s$ models the soft phase as hyperelastic, capturing the entropic elasticity and limited extensibility of polymer chains. As the hard phase breaks down, the hard segments become intermixed within the soft phase, leading to an elastomer-like response. Thus, we model the deviatoric part of mechanism $h1$ the same as the deviatoric part of mechanism $s$.

The contribution of each mechanism to the response at every material point is governed by the volume fractions of the hard and soft phases, which are calculated from MD simulation results. $\phi_{h}$ is the hard phase volume fraction and $\phi_{s} = (1 - \phi_{h})$ is the soft phase volume fraction. Hard phase breakdown is captured by volume fraction shifting from mechanism $h0$ to $h1$. This is captured by the state variable $\xi_{h0}$. We model breakdown as $\xi_{h0}$ monotonically decreasing from one to a value greater than or equal to zero. The broken-down phase (mechanism $h1$) is modeled separately from the initial soft phase (mechanism $s$) to conceptually indicate the conservation of mass during the breakdown process. In addition, the separate phase allows us to model the breakdown process having an associated decrease in density. We propose that during hard phase breakdown, the tightly aggregated glassy hard phase becomes intermixed with the elastomeric soft phase, resulting in density reduction. Notably, the methylene diphenyl diisocyanate (MDI) precursor, which forms the hard phase, has a density of approximately $\sim$1.2 $\text{g}/\text{cm}^{3}$. In contrast, the Versalink polytetrahydrofuran (polyTHF) chains comprising the soft segment have a density of $\sim$0.98 $\text{g}/\text{cm}^{3}$. We posit that the hard phase within PU closely matches the density of the MDI precursor, and as it mixes with the soft phase, its density approaches that of polyTHF.

The total strain energy density for each material point is the strain energy density of each mechanism weighted by $\phi_{h}$ and $\xi_{h0}$:
\begin{equation}
\label{eqn:total_free_energy}
    W = \phi_{h}\xi_{h0}(W_{h0^{\alpha}} + W_{h0^{\beta}}) + \phi_{h}(1 - \xi_{h0})W_{h1} + (1 - \phi_{h})W_{s}
\end{equation}
For each mechanism, we model the deviatoric and volumetric parts separately. The strain energy density for each mechanism is given by,
\begin{equation}
\label{eqn:free_energy_vol_dev}
W_{x} = W_{x}^{'} + W_{x}^{vol},
\end{equation}
where $W_{x}^{'}$ is the deviatoric part and $W_{x}^{vol}$ is the volumetric part with $x \in \{h0^{\alpha}, h0^{\beta}, h1, s \}$.

We construct our model using finite deformation measures because polyurea can undergo large deformations before failure. We assume that the individual mechanisms are subjected to the same deformation gradient $\bm{F}$. We can obtain the stress function for each mechanism by taking the partial derivative of each strain energy density function with respect to the appropriate strain measures based on $\bm{F}$. Specifically, we use the derivation for calculating Cauchy stress from strain energy found in \citet{bower_applied_2009}. From Equation \ref{eqn:total_free_energy}, the total Cauchy stress is given by the stress of each mechanism weighted by $\phi_{h}$ and $\xi_{h0}$:
\begin{equation} 
    \label{eqn:total_stress}
    \bm{T} = \xi_{h0}\phi_{h}(\bm{T}_{h0^{\alpha}} + \bm{T}_{h0^{\beta}}) + (1 - \xi_{h0})\phi_{h}\bm{T}_{h1} + (1 - \phi_{h})\bm{T}_{s}.
\end{equation}
The Cauchy stress is also split into deviatoric and volumetric parts, $\bm{T}_{x} = \bm{T}_{x}^{'} + \bm{T}_{x}^{vol}$, where $\bm{T}_{x}^{'}$ is the deviatoric stress and $\bm{T}_{x}^{vol}$ is the volumetric stress with $x \in \{h0^{\alpha}, h0^{\beta}, h1, s \}$.

\subsection{Volumetric response}
\subsubsection{Volumetric response of $h0$ and $h1$}
We use the Jacobian of the deformation gradient, $J = det(\bm{F})$, as a measure of volume change for each material point. The transition from a tightly aggregated glassy phase (mechanism $h0$) into an elastomeric phase (mechanism $h1$) results in density reduction during breakdown. $J$ is split into the part due to the density change $J^{h0 \rightarrow h1}$, and the part due to elastic deformation $J^{e}$, where
\begin{equation}
\label{eqn:JeJdens}
    J = J^{e}J^{h0 \rightarrow h1}.
\end{equation}
$J^{h0 \rightarrow h1}$ is related to the density of the initial hard phase ($\rho_{h0^{\alpha}} = \rho_{h0^{\beta}} = \rho_{h0}$), the density of the broken-down hard phase ($\rho_{h1}$), and the state variable controlling breakdown ($\xi_{h0}$) by
\begin{equation}
\label{eqn:Jdens}
    J^{h0 \rightarrow h1} = \dfrac{\rho_{h0}}{\left(\rho_{h1} - \rho_{h0}\right)\left(1 - \xi_{h0}\right) + \rho_{h0}}.
\end{equation}
We specify that only $J^{e}$ contributes to volumetric strain energy for mechanisms $h0^{\alpha}$, $h0^{\beta}$, and $h1$. We use the following volumetric strain energy function:
\begin{equation}
    \label{eqn:hard_vol_energy}
    W_{x}^{vol} = \dfrac{k_{x}}{2}(J^{e} - 1)^{2},
\end{equation}
where $k_{x}$ is the bulk modulus with $x \in \{h0^{\alpha}, h0^{\beta}, h1\}$.  The volumetric stress is then given by,
\begin{equation}
    \bm{T}_{x}^{vol} = k_{x}(J^{e} - 1)\bm{1},
\end{equation}
where $\bm{1}$ is the identity matrix and $x \in \{h0^{\alpha}, h0^{\beta}, h1\}$.

\subsubsection{Volumetric response of mechanism $s$}
The volumetric response of the soft phase (mechanism $s$) is a function of $J$. Using the same strain energy function as Equation \ref{eqn:hard_vol_energy}, the strain energy for mechanism $s$ is given by, \begin{equation}
   \label{eqn:soft_vol_energy}
    W_{s}^{vol} = \dfrac{k_{s}}{2}(J - 1)^{2},
\end{equation}
where $k_{s}$ is the bulk modulus. The volumetric stress is then given by
\begin{equation}
    \bm{T}_{s}^{vol} = k_{s}(J - 1)\bm{1}.
\end{equation}

\subsection{Deviatoric response}
\subsubsection{Deviatoric response of mechanism $h0$}
Mechanism $h0^{\alpha}$ represents the intermolecular resistance to deformation and subsequent yielding of the hard phase. The following elastic-viscoplastic model takes its main concepts from a glassy polymer model developed by \citet{mulliken_mechanics_2006}. The total deformation results from both elastic and plastic parts, which are modeled by multiplicative decomposition of the deformation gradient, $\bm{F} = \bm{F}_{h0^{\alpha}}^{e}\bm{F}_{h0^{\alpha}}^{p}$, where $\bm{F}_{h0^{\alpha}}^{e}$ is the elastic deformation gradient and $\bm{F}_{h0^{\alpha}}^{p}$ is the plastic deformation gradient. The deformation rates are described by the velocity gradient tensor, $\bm{L} = \bm{\dot{F}}\bm{F}^{-1}$. We decompose the velocity gradient of mechanism $h0^{\alpha}$ into its elastic and plastic parts:
\begin{equation}
\bm{L}_{h0^{\alpha}} = \bm{L}_{h0^{\alpha}}^{e} + \bm{F}_{h0^{\alpha}}^{e}\bm{L}_{h0^{\alpha}}^{p}\bm{F}_{h0^{\alpha}}^{e-1} = \bm{L}_{h0^{\alpha}}^{e} + \widetilde{\bm{L}}_{h0^{\alpha}}^{p},
\end{equation}
\begin{equation}
\bm{L}_{h0^{\alpha}}^{e} = \bm{\dot{F}}_{h0^{\alpha}}^{e}\bm{F}_{h0^{\alpha}}^{e-1},
\end{equation}
and
\begin{equation}
\bm{L}_{h0^{\alpha}}^{p} = \bm{\dot{F}}_{h0^{\alpha}}^{p}\bm{F}_{h0^{\alpha}}^{p-1}.
\end{equation}
$\bm{L}_{h0^{\alpha}}^{p}$ is the plastic velocity gradient in the reference configuration and $\widetilde{\bm{L}}_{h0^{\alpha}}^{p}$ is the plastic velocity gradient in the deformed configuration. The velocity gradient tensor can be decomposed into symmetric and skew-symmetric parts. For the plastic velocity gradient in the deformed configuration, 
\begin{equation}
\widetilde{\bm{L}}_{h0^{\alpha}}^{p} = \widetilde{\bm{D}}_{h0^{\alpha}}^{p} + \widetilde{\bm{W}}_{h0^{\alpha}}^{p},
\end{equation}
where $\widetilde{\bm{D}}_{h0^{\alpha}}^{p}$ is the symmetric part called the rate of deformation tensor and $\widetilde{\bm{W}}_{h0^{\alpha}}^{p}$ is the skew-symmetric part called the spin tensor. We make the typical assumption that plastic flow is irrotational, thus $\widetilde{\bm{W}}_{h0^{\alpha}}^{p} = 0$. The plastic deformation gradient of mechanism $h0^{\alpha}$ is updated by
\begin{equation}
\bm{\dot{F}}_{h0^{\alpha}}^{p} = \widetilde{\bm{L}}_{h0^{\alpha}}^{p}{\bm{F}}_{h0^{\alpha}}^{p} = \bm{F}_{h0^{\alpha}}^{e-1}\widetilde{\bm{D}}_{h0^{\alpha}}^{p}\bm{F}.
\end{equation}
We relate the rate of plastic deformation to deviatoric stress in a thermally activated yield process by the following constitutive equations:
\begin{equation}
    \label{eqn:plastic_flow_1}
    \widetilde{\bm{D}}^{p}_{h0^{\alpha}} = \dot{\gamma}^{p}_{h0^{\alpha}}\frac{\bm{T}_{h0^{\alpha}}^{'}}{|| \bm{T}_{h0^{\alpha}}^{'}||},
\end{equation}
and
\begin{equation}
    \label{eqn:plastic_flow_2}
    \dot{\gamma}^{p}_{h0^{\alpha}} = \dot{\gamma}^{0}_{h0^{\alpha}} \exp\Bigg[ -\frac{\Delta G_{h0^{\alpha}}}{k_b\theta} \Bigg(1 - \frac{\tau_{h0^{\alpha}}}{s^{0}_{h0^{\alpha}}} \Bigg) \Bigg].
\end{equation}
$\bm{T}_{h0^{\alpha}}^{'}$ is the deviatoric stress of mechanism $h0^{\alpha}$, and $|| \bm{T}_{h0^{\alpha}}^{'} ||$ is the Frobenius norm of the deviatoric stress. $\tau_{h0^{\alpha}} = \big(\frac{1}{2} \bm{T}_{h0^{\alpha}}^{'}\bm{T}_{h0^{\alpha}}^{'}\big)^{\frac{1}{2}}$ is the scalar equivalent shear stress. $s^{0}_{h0^{\alpha}}$ is the athermal shear strength, $\Delta G_{h0^{\alpha}}$ is activation energy, $\dot{\gamma}^{0}_{h0^{\alpha}}$ is the attempt frequency of shear transition, $k_b$ is Boltzmann's constant, and $\theta$ is the temperature. The athermal shear strength is related to shear modulus and Poisson's ratio by,
\begin{equation}
\label{eqn:athermal_shear_strength}
s^{0}_{h0^{\alpha}} = \frac{0.077 \mu_{h0^{\alpha}}}{1 - \nu_{h0^{\alpha}}}, 
\end{equation}
where $\mu_{h0^{\alpha}}$ is shear modulus and $\nu_{h0^{\alpha}}$ is Poisson's ratio \citep{argon_theory_1973}. The attempt frequency of shear transition, $\dot{\gamma}^{0}_{h0^{\alpha}}$, and the activation energy $\Delta G_{h0^{\alpha}}$ are material properties of the hard phase.

The stress is a function of strain measures from the elastic deformation gradient, $\bm{F}^{e}_{h0^{\alpha}}$. From the elastic deformation gradient, we obtain the elastic deviatoric Almansi strain,
\begin{equation}
    \bm{e}_{h0^{\alpha}}^{e'} = \dfrac{1}{2}(\bm{1} - \bm{B}_{h0^{\alpha}}^{e'-1}),
\end{equation}
where $\bm{B}_{h0^{\alpha}}^{e'}=J^{-\frac{2}{3}}\bm{F}_{h0^{\alpha}}^{e}\bm{F}_{h0^{\alpha}}^{eT}$ is the deviatoric left elastic Cauchy-Green tensor. Almansi strain and Cauchy stress are energetic conjugates, so the strain energy density for the deviatoric part of mechanism $h0^{\alpha}$ is given by,
\begin{equation}
    W_{h0^{\alpha}}^{'} = \bm{T}_{h0^{\alpha}}^{'} : \bm{e}_{h0^{\alpha}}^{e'}.
\end{equation}
The deviatoric linear-elastic stress is given by
\begin{equation}
\bm{T}_{h0^{\alpha}}^{'} = 2\mu_{h0^{\alpha}}\bm{e}_{h0^{\alpha}}^{e'},
\end{equation}
where $\mu_{h0^{\alpha}}$ is the shear modulus.

In parallel, mechanism $h0^{\beta}$ captures the entropic resistance to chain alignment within the hard phase. We model this with a neo-Hookean hyperelastic element. We obtain the deviatoric neo-Hookean stress by substituting deviatoric strain measures into the neo-Hookean strain energy density function,
\begin{equation}
    \label{eqn:neohookean_energy_h0beta}
    W^{'}_{h0^{\beta}} = \frac{\mu_{h0^{\beta}}}{2}({I}^{'}_{1} - 3),
\end{equation}
\noindent where $\mu_{h0^{\beta}}$ is the shear modulus, and
\begin{equation}
    \label{eqn:deviatoricLeftCauchyGreen}
    {I}^{'}_{1} = tr(\bm{B}^{'}) = tr(J^{-\frac{2}{3}}\bm{F}\bm{F}^{T}) 
\end{equation}
is the first invariant of the deviatoric left Cauchy-Green tensor.
The deviatoric stress is given by,
\begin{equation}
    \label{eqn:neohookean_stress_h0beta}
    \bm{T}_{h0^{\beta}}^{'} = \mu_{h0^{\beta}}\Big(\frac{1}{J^{5/3}}\bm{B} - \frac{1}{3J} {I}^{'}_{1}\bm{1} \Big),
\end{equation}
where $\bm{B} = \bm{F}\bm{F}^T$ is the left Cauchy-Green tensor. 

\subsubsection{Deviatoric response of mechanisms $s$ and $h1$}

Mechanism $s$ represents the elastomeric soft phase. The deviatoric part is modeled with a Gent hyperelastic element. 
As the hard phase breaks down from deformation, the hard segments become occluded within the soft phase. We represent the broken-down phase as mechanism $h1$. Because mechanism $h1$ represents fully occluded hard segments, we model the deviatoric part of mechanism $h1$ with the same Gent model as mechanism $s$. In the following model description, the subscript $x$ denotes both mechanism $s$ and $h1$. 

To obtain a deviatoric Gent stress, we substitute the elastic deviatoric first invariant of the left Cauchy-Green strain tensor $I_{1 x}^{e'}$ (eq. \ref{eqn:deviatoricLeftCauchyGreen}), into the Gent strain energy function:
\begin{equation}
    W^{'}_{x} = -\frac{\mu_{x}(I^{m}_{x}-3)}{2}\ln\Bigg(1 - \frac{I^{e'}_{1x} - 3}{I^{m}_{x} - 3}\Bigg),
\end{equation}
where $\mu_{x}$ is the shear modulus, and $I^{m}_{x}$ is the limiting stretch with $x \in \{s,h1\}$.
The strain energy function has a singularity when $I^{e'}_{1x}$ reaches $I^{m}_{x}$ causing a large increase in stress. The limiting stretch is related to the length of polymer chains by $I^{m}_{x} \approx 3N$, where $N$ is the number of Kuhn segments \citep{beatty_average-stretch_2003,rickaby_comparison_2015}. The deviatoric stress is given by
\begin{equation}
\mathbf{T}_{x}^{'} = \dfrac{1}{J^{5/3}}\Bigg(\frac{\mu_{x} (I^{m}_{x}-3)}{I^{m}_{x} - I_{1x}^{e'}} \Bigg)\mathbf{B} - \frac{1}{3J}I_{1x}^{e'}\Bigg(\frac{\mu_{x} (I^{m}_{x}-3)}{I^{m}_{x} - I_{1x}^{e'}} \Bigg)\mathbf{1},
\end{equation}
with $x \in \{s,h1\}$. $\bm{B} = \bm{F}\bm{F}^{T}$ is the left Cauchy-Green deformation tensor.

\subsection{Hard phase breakdown evolution}
During loading, the hard segments disaggregate and become intermixed
with the soft segments. In this model, the amount of hard phase breakdown is controlled by the state variable $\xi_{h0}$. We propose that this process is dissipative and does not cause an increase in strain energy density. If the following inequality is satisfied, there will be no increase in strain energy density as $\xi_{h0}$ evolves:
\begin{equation}
\label{eqn:dissipation_ineq1}
    \frac{\partial W}{\partial \xi_{h0}} \dot{\xi}_{h0} \leq 0.
\end{equation}
Substituting equations \ref{eqn:total_free_energy} and \ref{eqn:free_energy_vol_dev} into inequality \ref{eqn:dissipation_ineq1}, and noting that the deviatoric portions of each free energy are not functions of the hard phase breakdown, gives the following inequality:
\begin{equation}
\label{eqn:dissipation_ineq2}
\Bigg\{  \phi_{h}  \Big(W_{h0^{\alpha}}^{'} + W_{h0^{\beta}}^{'} - W_{h1}^{'} \Big)
+ \frac{\partial}{\partial \xi_{h0}}\Big[ \phi_{h} \xi_{h0}(W_{h0^{\alpha}}^{vol} + W_{h0^{\beta}}^{vol}) + \phi_{h}(1 - \xi_{h0})(W_{h1}^{vol}) + (1- \phi_{h})(W_{s}^{vol})     \Big] \Bigg\} \dot{\xi}_{h0} \leq 0
\end{equation}
We prescribe ${\xi}_{h0}$ to be monotonically decreasing, that is $\dot{\xi}_{h0}\leq 0$. We satisfy inequality \ref{eqn:dissipation_ineq2} by considering the deviatoric and volumetric parts separately. That is,
\begin{equation}
    \label{eqn:dissipation_ineq3}
    W_{h0^{\alpha}}^{'} + W_{h0^{\beta}}^{'} - W_{h1^{\alpha}}^{'} \geq 0,
\end{equation}
and
\begin{equation}
\label{eqn:dissipation_ineq4}
\frac{\partial}{\partial \xi_{h0}}\Big[  \phi_{h}\xi_{h0}(W_{h0^{\alpha}}^{vol} + W_{h0^{\beta}}^{vol}) + \phi_{h}(1 - \xi_{h0})(W_{h1}^{vol})  + (1 - \phi_{h})(W_{s}^{vol})  \Big] \geq 0.
\end{equation}
To satisfy the condition on the deviatoric part, inequality \ref{eqn:dissipation_ineq3}, we prescribe that breakdown only occurs when the deviatoric strain energy of mechanism $h0$ is greater than or equal to that of mechanism $h1$. For the volumetric part, if we substitute equations \ref{eqn:JeJdens}, \ref{eqn:Jdens}, and \ref{eqn:hard_vol_energy} into inequality \ref{eqn:dissipation_ineq4}, we find that the inequality is satisfied if $J^{e} \geq 1$ when the bulk moduli and density of the mechanism $h0$ are prescribed to be greater or equal to that of mechanism $h1$ (Supplementary Materials Section S2).
\begin{equation}
\left(k_{h0^{\alpha}} + k_{h0^{\beta}}\right) \geq  k_{h1},
\end{equation}
and
\begin{equation}
\rho_{h0} \geq \rho_{h1}.
\end{equation}
The physical interpretation of the $J^{e} \geq 1$ condition for breakdown is that loading-induced volumetric expansion is required for the $h0$ phase to transition into the less dense $h1$ phase. Otherwise, the density change could cause a non-physical increase in strain energy density.

We propose that plastic distortion of the hard phase is required for the hard segments to break apart and become intermixed with the soft phase. We model hard phase breakdown as driven by plastic deformation of the hard phase. The rate of hard phase breakdown $\dot{\xi}_{h0}$ evolves with plastic strain rate $\dot{\gamma}^{p}_{h0^{\alpha}}$ as follows:
\begin{equation}
    \label{eqn:hard_phase_breakdown_1}
    \dot{\xi}_{h0} = C \Bigg( \dfrac{\dot{\gamma}^{p}_{h0^{\alpha}}}{\dot{\xi}_{Base}} \Bigg) ^{m}
    \begin{cases} 
      1, &  \left(W_{h0^{\alpha}}^{'} + W_{h0^{\beta}}^{'}\right) \geq \left(W_{h1}^{'} \right)  \text{  and  } J^{e} \geq 1 \\  
      0, & \text{else}
    \end{cases}.
\end{equation}
The parameter $m$ allows plastic flow to drive hard phase breakdown in a non-linear way. $\dot{\xi}_{Base}$ is the resistance to breakdown and is included for dimensional consistency. $C/(\dot{\xi}_{Base})^{m}$ and $m$ are the material fitting parameters. The conditional expression ensures that $\xi_{h0}$ evolution does not result in an increase in strain energy. 

\section{Results and discussion}
\subsection{Molecular dynamics simulation and finite element RVE generation results}
\label{sec:results_md}

\begin{figure}[htb!]
	\centering
    \includegraphics[scale=0.75]{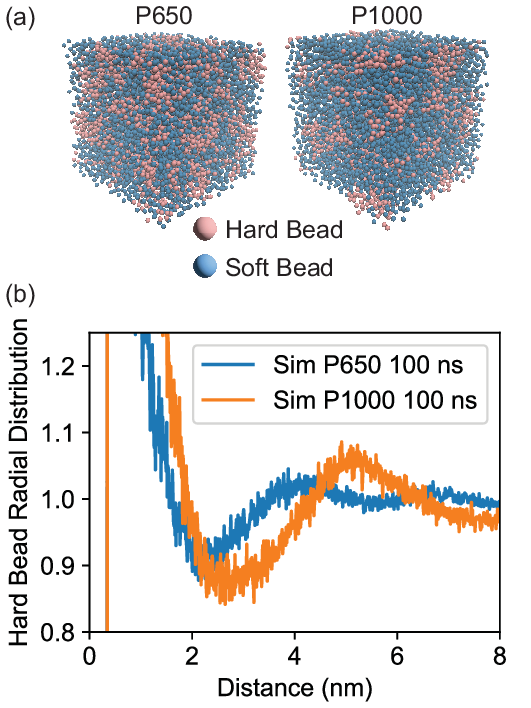}
	\caption{(a) Rendered snapshots of equilibrated P650 and P1000 coarse-grained molecular dynamics simulations. Pink beads represent the hard segments, while blue beads represent the soft segments. (b) Radial distributions of hard beads from P650 (blue line) and P1000 (orange line) simulations, with peaks at 4.2 nm and 5.1 nm, respectively.}
	\label{FIG:CGMD_Results_1}
\end{figure}

The equilibrated configurations of coarse-grained molecular dynamics P650 and P1000 simulations are shown in Figure \ref{FIG:CGMD_Results_1}a. The snapshots were captured after 100 ns of equilibration at 300 K and 1 atm. In Figure \ref{FIG:CGMD_Results_1}a, it is evident that the hard segments, shown as pink beads, have aggregated into domains. Moreover, the domains within the P1000 system appear more dispersed than P650 due to P1000 having longer soft segments. To quantify hard domain dispersion, we generated the radial distribution function of the hard beads (Figure \ref{FIG:CGMD_Results_1}b). We take the peak occurring at $\sim$4.2 nm for P650 and $\sim$5.1 nm for P1000 to be the hard phase interdomain spacings. The simulated interdomain spacings are somewhat lower than that obtained from small-angle X-ray scattering experiments of $\sim$6.4 nm for P650 and $\sim$7.2 nm for P1000 \citep{castagna_role_2012}. During our simulated equilibration process, we found that the increase in hard domain spacing plateaus after $\sim$50 ns. The disparity between the domain spacings from our simulations versus experiments could be due to continuing phase segregation occurring at significantly longer timescales that cannot be readily modeled. In addition, there may be discrepancies between the simulated polymer and actual polymer compositions. For instance, while the simulated polymer has monodisperse soft segments and degree of polymerization, the actual polymer likely has polydispersity. Overall, the simulated polymer provides a close approximation to the actual polymer, with P1000 displaying a greater interdomain spacing compared to P650. 

To obtain volume fraction information for finite element analysis, we generated an RVE grid based on the results of CGMD simulations, as described in Section \ref{sec:methods_cgmd_2_fe_rve}. Supplementary materials Figure S3 displays snapshots of the resulting RVEs. We verified the RVE generation procedure by computing the average hard phase volume fraction over the RVE. We obtained a value of 0.353 for P650 and 0.230 for P1000. These values closely match the hard-to-soft bead number and mass fractions of 0.381 for P650 and 0.258 for P1000 in the GCMD simulations, confirming that the RVE generation process conserves the overall weight fraction of the beads. 

\subsection{Constitutive model parameter identification}
\label{sec:results_parameter_id}
\begin{figure}[htb!]
	\centering
		\includegraphics[scale=0.75]{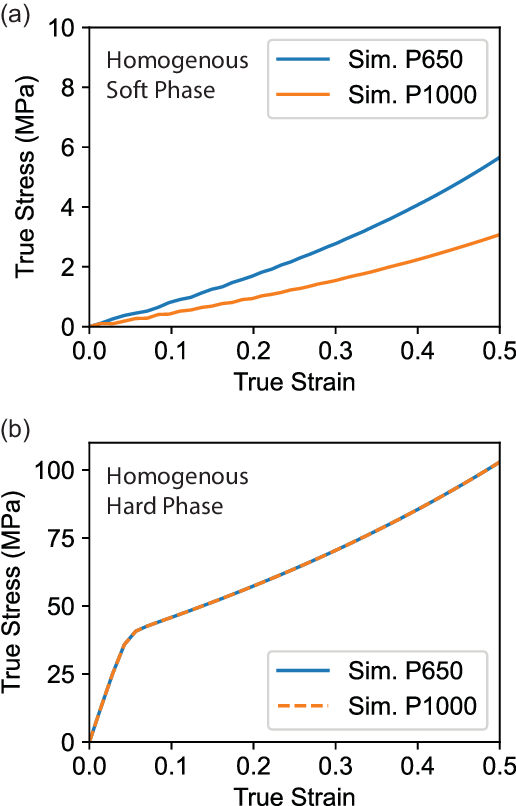}
	\caption{True stress vs. true strain data showing the behavior of homogeneous finite element simulations undergoing monotonic loading at a strain rate of 0.1 $\text{s}^{\text{-1}}$. For Subfigure (a), the simulation is only of the soft phase, while (b) is only of the hard phase, with no hard phase breakdown. The blue curves represent the simulation results using P650 material properties. Conversely, the orange curves (dashed in (b) for clarity) represent the results of simulations featuring P1000 material properties.}
	\label{FIG:parameter_id}
\end{figure}

We used physically inspired quantities to determine the soft phase material properties. The shear modulus of mechanism $s$ was approximated based on the chain density of the P650 and P1000 soft segments, with molecular weights of 973 g/mol and 1658 g/mol, respectively. The formula $\mu_{s} = (\rho_{s}R\theta)/M_{c}$ was used to calculate the shear modulus using the density of the soft phase ($\rho_{s} \approx 1$ $\text{g}/\text{cm}^{3}$), universal gas constant (R), simulation temperature ($\theta$ = 293 K), and molecular weight of each chain ($M_{c}$). Moreover, we considered the chain segment length to determine the limiting stretch within the Gent model. It has been shown that the limiting stretch of $\text{I}_1$, denoted as $\text{I}^{m}_{s}$, is equal to three times the number of Kuhn segments in the chain segment \citep{beatty_average-stretch_2003,rickaby_comparison_2015}. In order to determine the number of Kuhn segments, we assumed that the characteristic ratio and bond angles of the polytetrahydrofuran soft segments are similar to those of polyethylene glycol, which has a characteristic ratio of 6.9 and a bond angle of 70 degrees. Based on these values, we calculated the ratio between the number of bonds and Kuhn segments to be $\sim10.3:1$. In our molecular dynamics simulation, each coarse-grained bead represents four atoms. Thus, for our simulated P650, which has 52 beads, we estimated around five Kuhn segments. For the P1000 simulation, which has 92 beads, we estimated around 9 Kuhn segments. Multiplying by three, we approximated $\text{I}^{m}_{s}$ to be 15 for P650 and 27 for P1000. Given that the soft phase exhibits elastomeric behavior and is relatively incompressible, we assumed a Poisson's ratio of approximately 0.49 when calculating the bulk modulus using the shear modulus. Figure \ref{FIG:parameter_id}a shows the stress-strain curve of the soft phase stress-strain response for a homogeneous material with P650 and P1000 material properties. The broken-down hard phase parameters are taken to be the same as the soft phase following the same logic since we consider the broken-down hard phase as previously occluded soft segments.

To determine the material parameters for the hard phase (mechanisms $h0^{\alpha}$ and $h0^{\beta}$), we fitted the P650 RVE response to experimental stress-strain data and then used the same parameters for P1000. Since the material model suggests that the intermolecular forces of the hard phase, represented as mechanism $h0^{\alpha}$, dominate the initial linear elastic region of the stress-strain curve, we first fitted the shear modulus ($\mu_{h0^{\alpha}}$) of mechanism $h0^{\alpha}$ to this region. Similarly, because mechanism $h0^{\beta}$ represents the resistance to chain orientation during larger deformations, we fitted the shear modulus $\mu_{h0^{\beta}}$ to the post-yield region of the stress-strain curve. Assuming a typical Poisson's ratio of 0.35 for the glassy hard phase, we calculated the bulk moduli for the hard phase, $k_{h0^{\alpha}}$ and $k_{h0^{\beta}}$, using the Poisson's ratio and the shear moduli. Moreover, the athermal shear strength for mechanism $h0^{\alpha}$ $\left(s_{h0^{\alpha}}^{0}\right)$ was determined through equation \ref{eqn:athermal_shear_strength}. To determine the viscoplastic flow parameters, $\dot{\gamma}_{h0^{\alpha}}^{0}$ and $\Delta G_{h0^{\alpha}}$, we fitted equation \ref{eqn:plastic_flow_2} to the yield stress data obtained from P650 experiments at loading rates of 0.001 to 0.1 $\text{s}^{-1}$. As polyurea is a composite of hard and soft phases, and the viscoplastic flow parameters only describe the hard phase, we scaled the experimentally observed yield stress by the weight fraction of hard segments to find these. Figure \ref{FIG:parameter_id}b shows the hard phase response (mechanisms $h0^{\alpha}$ and $h0^{\beta}$) for P650 and P1000 material properties. Notably, the hard phase stress-strain curves are the same for P650 and P1000 due to identical material parameters.

We obtained the hard phase breakdown parameters by fitting the simulated P650 RVE to experimental data. The pre-power parameter $C$ was set to -1 $\text{s}^{-1}$, so the volume fraction of the hard phase decreases with increasing plastic strain. As the viscoplastic flow rule for the hard phase and the hard phase breakdown process are the only rate-dependent elements in the model, the power parameter $m$ was fitted to the observed difference in flow stress for P650 experiments at different rates. $\dot{\xi}_{Base}$ was set to 1 $\text{s}^{-1}$ and included only for dimensional consistency. The material parameters for finite element simulations of the P650 and P1000 RVE are summarized in Table \ref{tbl:material_parameters}.

\begin{table}[htb!]
\centering
\caption{Model parameters for finite element simulation of P650 and P1000.}
\label{tbl:material_parameters}
\begin{tabular}{p{0.275\textwidth}p{0.275\textwidth}p{0.275\textwidth}}
\toprule
    & \textbf{Sim. P650} & \textbf{Sim. P1000}  \\
\midrule
    \textbf{Mechanism $h0^{\alpha}$} \\
    $\mu_{h0^{\alpha}}$   &  $300$ ($\text{MPa}$)   &  $300$  ($\text{MPa}$)  \\
    $k_{h0^{\alpha}}$ &  $900$ ($\text{MPa}$)   &  $900$ ($\text{MPa}$)  \\
    $\rho_{h0^{\alpha}}$  &  $1.2$ ($\text{g}/\text{cm}^{\text{3}}$) &  $1.2$ ($\text{g}/\text{cm}^{\text{3}}$)  \\
    $s_{h0^{\alpha}}^0$  &  $35.54$ ($\text{MPa}$)  &  $35.54$ ($\text{MPa}$)  \\
    $\dot{\gamma}^{0}_{h0^{\alpha}}$ & $52.96$ ($\text{1/s}$) & $52.96$ ($\text{1/s}$) \\
    $\Delta G_{h0^{\alpha}}$ & $5.905 \times 10^{-20}$  ($\text{J}$) & $5.905 \times 10^{-20}$ ($\text{J}$) \\
    $\theta$ & $293$ ($\text{K}$)  & $293$ ($\text{K}$)  \\
  \midrule
    \textbf{Mechanism $h0^{\beta}$} \\
    $\mu_{h0^{\beta}}$ & $35$ ($\text{MPa}$) & $35$ ($\text{MPa}$)  \\
    $k_{h0^{\beta}}$ & $105$ ($\text{MPa}$)  & $105$ ($\text{MPa}$) \\
    $\rho_{h0^{\beta}}$  &  $1.2$ ($\text{g}/\text{cm}^{\text{3}}$) &  $1.2$ ($\text{g}/\text{cm}^{\text{3}}$)  \\
  \midrule
    \textbf{Mechanism $s$} \\
    $\mu_{s}$ & $2.55$ ($\text{MPa}$) & $1.44$ ($\text{MPa}$) \\
    $I_{s}^{m}$  & $15$  & $27$  \\
    $k_{s}$  &  $126.65$ ($\text{MPa}$)   &  $71.52$ ($\text{MPa}$)    \\
  \midrule
  \textbf{Mechanism $h1$}\\
    $\mu_{h1}$   & $2.55$ ($\text{MPa}$)    &  $1.44$ ($\text{MPa}$)    \\
    $I_{h1}^{m}$     & $15$    &  $27$    \\
    $k_{h1}$    &  $126.65$ ($\text{MPa}$)   &  $71.52$ ($\text{MPa}$)    \\
    $\rho_{h1}$  &  $1.0$ ($\text{g}/\text{cm}^{\text{3}}$) &  $1.0$ ($\text{g}/\text{cm}^{\text{3}}$)  \\
   \midrule
  \textbf{Hard Phase Breakdown}  \\
    $C$   & $-1$ ($\text{s}^{\text{-1}}$) &  $-1$ ($\text{s}^{\text{-1}}$) \\
    $\dot{\xi}_{Base}$     &    $1$  ($\text{s}^{\text{-1}}$)    &  $1$  ($\text{s}^{\text{-1}}$)   \\
    $m$  &   $0.8$  &    $0.8$   \\
\bottomrule
\end{tabular}
\end{table}

\FloatBarrier

\subsection{Comparing finite element RVE simulation with experimental data}
\label{sec:results_fe_exp}
\begin{figure}[htb!]
	\centering
		\includegraphics[scale=0.75]{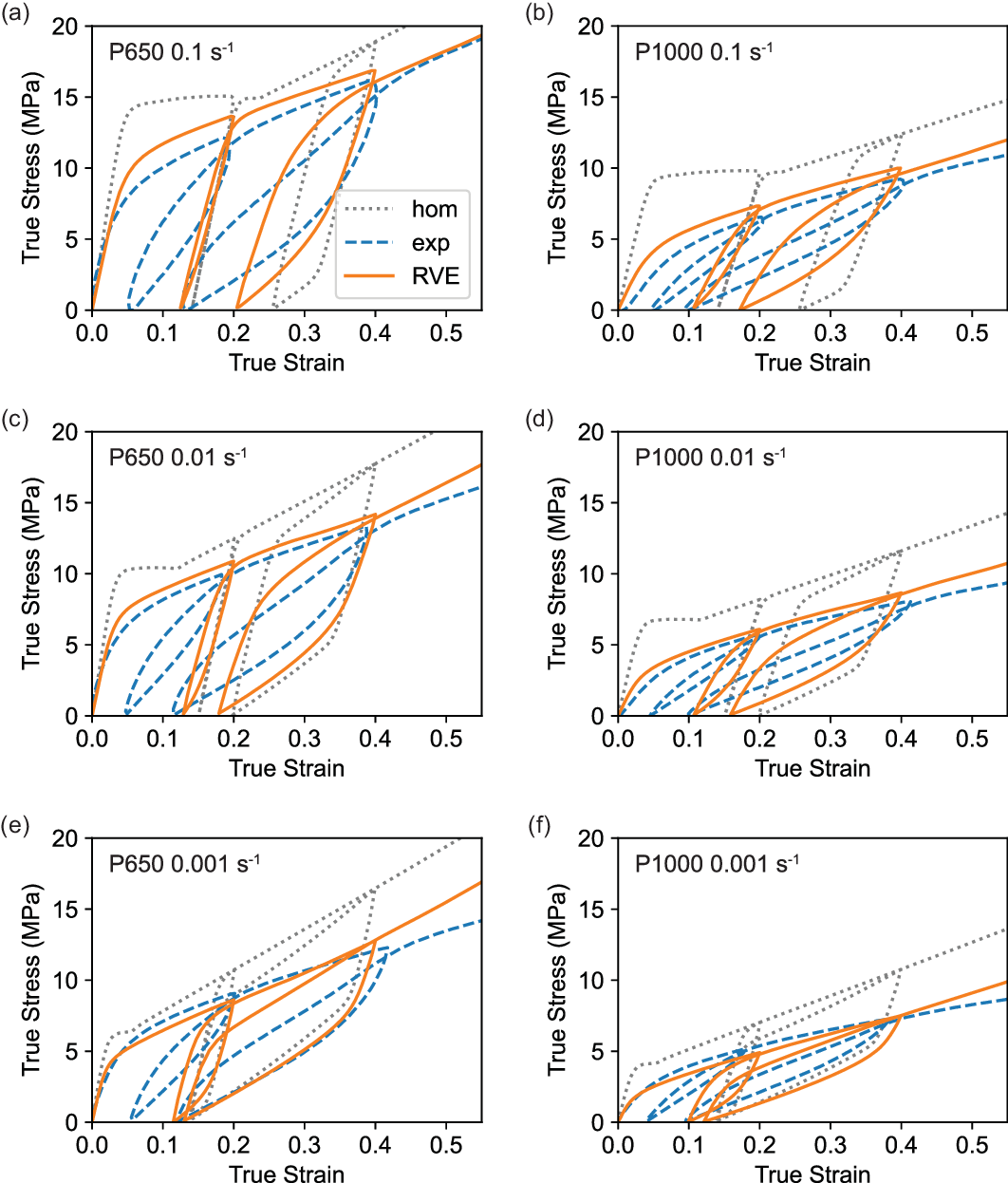}
	\caption{Cyclic true stress vs. true strain data from our simulations and experiments. The data includes the following materials and loading strain rates: (a) P650 and (b) P1000 at 0.1 $\text{s}^{\text{-1}}$. (c) P650 and (d) P1000 at 0.01 $\text{s}^{\text{-1}}$. (e) P650 and (f) P1000 at 0.001 $\text{s}^{\text{-1}}$. The dashed blue line corresponds to our experimental results. The solid orange line is from simulations using our heterogeneous RVEs. The gray dashed line is from homogeneous simulations, where the initial hard phase volume fraction matches the average from our RVEs, specifically 0.353 for P650 and 0.230 for P1000.}
	\label{FIG:cyclic_stress_strain}
\end{figure}

Figure \ref{FIG:cyclic_stress_strain} shows a comparison of the simulated polyurea RVE to low-rate (0.001 - 0.1 $\text{s}^{\text{-1}}$) tensile cyclic loading experiments. The experimental results for both P650 and P1000 exhibit a linear elastic loading region at low strains, followed by a yield-like response. During cyclic loading experiments, at specific strain intervals, the specimens were unloaded to zero force and then reloaded. We observed a combination of permanent set and elastic recovery in the material, as evident from residual strain when the force reaches zero. The unloading and reloading curves display considerable hysteresis, and there is another yield-like response upon reloading. Upon reloading, the post-yield flow stress exhibits a similar slope as the previous load cycle. The cyclic loading stress-strain response rejoining the monotonic curve is in accordance with previously conducted PU experiments \citep{rinaldi_microstructure_2011}. As the strain rate increases, both P650 and P1000 experiments exhibit an increase in yield stress. In addition, P650 has a higher yield stress than P1000 at a given strain rate. The slope of the flow stress after yield is steeper for P650 than for P1000. However, for each material, the overall slope of the flow stress is similar from rates of 0.001 to 0.1 $\text{s}^{-1}$. 

A comparison of the P650 and P1000 RVE simulations with experimental data and with homogeneous simulations of the same average volume fraction emphasizes the significance of microstructure in predicting stress-strain responses (Figure \ref{FIG:cyclic_stress_strain}). As explained in Section \ref{sec:results_parameter_id}, the viscoplastic flow parameters affecting the yield stress were fitted solely using equation \ref{eqn:plastic_flow_2} to P650 low-rate experimental data. Remarkably, the simulations capture the rate dependence of the yield stress, as well as the lower yield stress of P1000. The variation in yield stress between P650 and P1000 can be partially attributed to the difference in the overall average hard phase volume fraction. However, we found that the homogeneous simulations over-predicted the yield stress, especially for P1000. This result is expected in the homogeneous simulations since all mechanisms are loaded affinely, with each mechanism experiencing the same strains. In contrast, in the RVE microstructure, a certain amount of hard phase is occluded, and the strain is non-uniform. The greater influence of the soft phase results in lower stresses for the RVE when compared to the homogeneous simulations. 

The influence of the RVE microstructure is also evident in the flow stress after yield. Both P650 and P1000 RVE simulations closely align with experimental flow stress, with the slopes remaining relatively consistent across low strain rates. In addition, the RVE simulations show that P650 exhibits a steeper flow stress slope compared to P1000. In contrast, while the homogeneous simulation can capture a difference in flow stress slope due to the different soft phase material parameters used for P650 and P1000, the overall profile of the curve differs from both the RVE simulations and experiments. Post-yield, the homogeneous stress-strain profile stays relatively flat before increasing at larger strains.  The hard phase breakdown process directly affects this profile, with a reduced breakdown rate observed as the stress upturn takes place  (Supplementary Materials Figure S4). On the other hand, the heterogeneous RVE microstructure enables various hard phase clusters to yield at different times, facilitating a smooth transition between elastic loading and plastic flow.

\begin{figure}[htb!]
	\centering
		\includegraphics[scale=0.75]{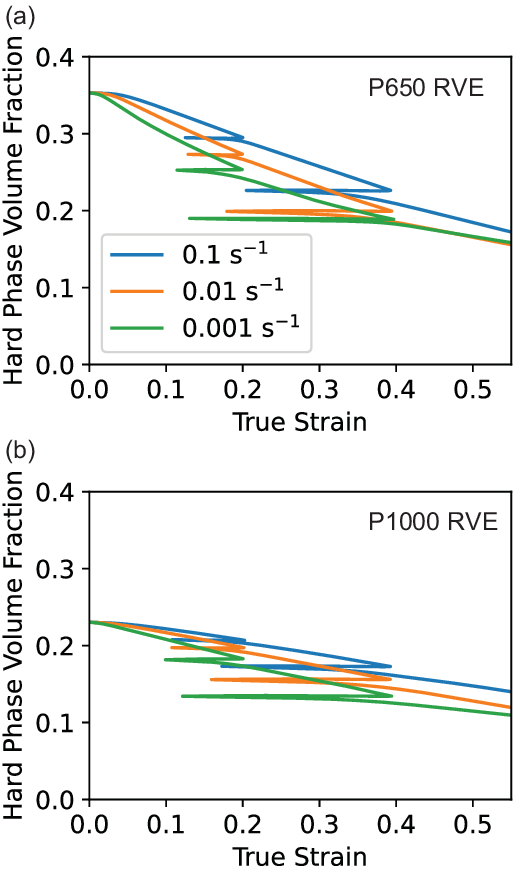}
        \caption{Hard phase volume fraction vs. strain for (a) P650 and (b) P1000 RVEs undergoing cyclic loading. The blue line is for simulations at a strain rate 0.1 $\text{s}^{\text{-1}}$, the orange line is for 0.01 $\text{s}^{\text{-1}}$, and the green line is for 0.001 $\text{s}^{\text{-1}}$. }
	\label{FIG:cyclic_volume_fraction}
\end{figure}

Similar to the experiments, during the loading and unloading cycles, the RVE simulations exhibit permanent deformation and hysteresis. Upon reloading, the RVE stress-strain yields and then rejoins the monotonic loading profile. When plotting the average hard phase volume fraction over the RVE (Figure \ref{FIG:cyclic_volume_fraction}), minimal breakdown is observed during unloading and reloading. If breakdown occurred during these cycles, the stress-strain curve would follow a different path compared to the monotonic curve, reflecting effects from a difference in material history. Compared to the experiments, the RVE simulation consistently under-predicts the strain at zero force after unloading. Additionally, at lower strains, the RVE simulation under-predicts the amount of hysteresis. One noteworthy difference between the RVE simulation and experiments is that, while the degree of strain recovery remains relatively constant across strain rates in experiments, the simulation predicts a higher degree of strain recovery upon unloading at lower strain rates than higher strain rates. The observed response stems from our rate-dependent hard phase breakdown process and the broken-down hard phase model (mechanism $h1$). As shown in Figure \ref{FIG:cyclic_volume_fraction}, lower rates exhibit more hard phase transition at a given strain due to the nonlinear relation between the rate of hard phase breakdown $\dot{\xi}_{h0}$ and the plastic flow rate $\dot{\gamma}_{h0^{\alpha}}^{p}$. We modeled the hard phase as breaking down into a stressed and fully elastic state, where the element's total deformation determines the stress. Thus, lower strain rate simulations with greater breakdown exhibit increased elastic recovery. In the actual material, the hard phase may break down into a relaxed state, which could cause the discrepancy. The impact of microstructure is evident when comparing the cyclic loading curves of the RVE and homogeneous simulations. In the homogeneous simulations, we observe almost no hysteresis at low strains, with nearly fully elastic unloading and reloading curves. A linear unloading curve appears at larger strains, followed by yielding and then linear reloading. The difference is that in the homogeneous simulation, the hard phase mechanism deforms and yields uniformly, whereas, in the RVE, the microstructure enables various hard phase clusters to yield at different times.

Further analysis of the average hard phase volume fraction as a function of strain for the low-rate cyclic experiment reveals a few key features (Figure \ref{FIG:cyclic_volume_fraction}). Initially, minimal breakdown occurs at small strains before yield because we model the breakdown process as being driven by plasticity. As the strain increases, we notice that the hard phase transition plateaus. The plateau is caused by the fact that the hard phase volume fraction cannot decrease below 0\% in combination with occluded hard phase clusters that experience minimal strain and, thus, exhibit minimal breakdown. The disparities in the plateau between P650 and P1000 can be attributed to differences in microstructure. Notably, the plateau in breakdown has also been observed in small-angle X-ray scattering studies. \citet{choi_microstructure_2012} conducted a study characterizing the degree of microphase separation of polyurea P1000 at increments of 100\% engineering strain to 400\% engineering strain. At 0\% strain, the degree of micro-phase separation was found to be 0.38. Interestingly, between 100\% and 300\% strain, the degree of phase separation remained relatively constant, ranging from 0.26 to 0.29. Although our simulations only go to ~70\% engineering strain, we believe that the observed plateau in our simulations is consistent with the experimental findings. 

\begin{figure}[htb!]
	\centering
		\includegraphics[scale=0.75]{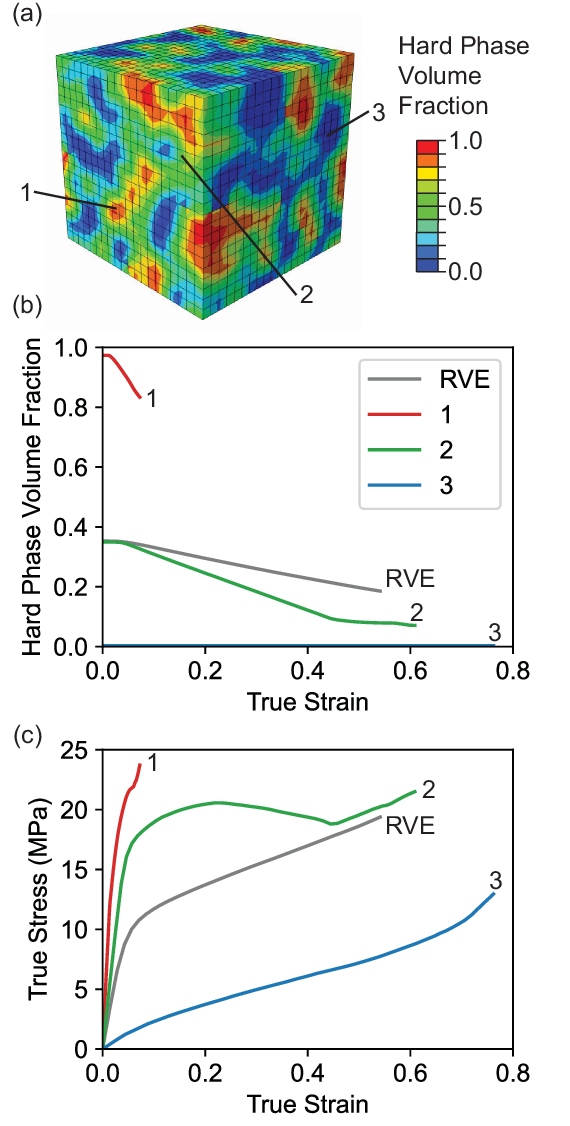}
	\caption{(a) The P650 RVE with the following elements labeled: 1 is an element within a hard domain, 2 corresponds to an element in the interfacial region, and 3 is an element in the soft phase. (b) Hard phase volume fraction vs. true strain, and (c) true stress vs. true strain for the elements above.}
	\label{FIG:RVE_Elements_Vf_Stress}
\end{figure}

Figure \ref{FIG:RVE_Elements_Vf_Stress} displays the hard phase volume fraction versus strain and stress-strain for particular elements of a P650 RVE. We present this figure to emphasize how strain localization influences distinct regions of the RVE microstructure. At the start of the simulation, element 1 represents an element in the hard phase, element 2 belongs to the interfacial region between the two phases, with an intermediate hard/soft phase content, and element 3 corresponds to an element in the soft phase. We deform the RVE monotonically to 0.55 strain. In contrast, element 1 only deforms to 0.074 true strain because the hard domain is relatively stiff compared to the surroundings. Figure \ref{FIG:RVE_Elements_Vf_Stress}b shows that element 1 has a relatively low amount of hard phase breakdown. This is attributed to the fact that the element stays in its elastic region for most of the simulation, as demonstrated in Figure \ref{FIG:RVE_Elements_Vf_Stress}c. Elements 2 and 3 experience substantially more deformation, with element 2 reaching 0.62 strain and element 3 reaching 0.77 strain. Despite the interfacial region (element 2) being relatively stiff before yield from the hard phase content, the region experiences considerable amounts of strain due to strain localization and amplification. Due to the large amount of strain, element 2 displays considerable hard phase breakdown, with the volume fraction decreasing from 0.35 to 0.07 (Figure \ref{FIG:RVE_Elements_Vf_Stress}b). For element 2, upon reaching 0.45 strain, the decrease in hard phase volume fraction slows, which corresponds to an increase in stress (Figure \ref{FIG:RVE_Elements_Vf_Stress}b). This is due to the material model's constraints, where the broken-down hard phase (mechanism $h1$) needs to have lower strain energy than the initial hard phase (mechanism $h0$) for breakdown to occur. The stress increase after 0.45 strain is associated with mechanism $h1$, and its strain energy inhibits further hard phase volume fraction decrease. 

\begin{figure}[htb!]
	\centering
    \includegraphics[scale=0.75]{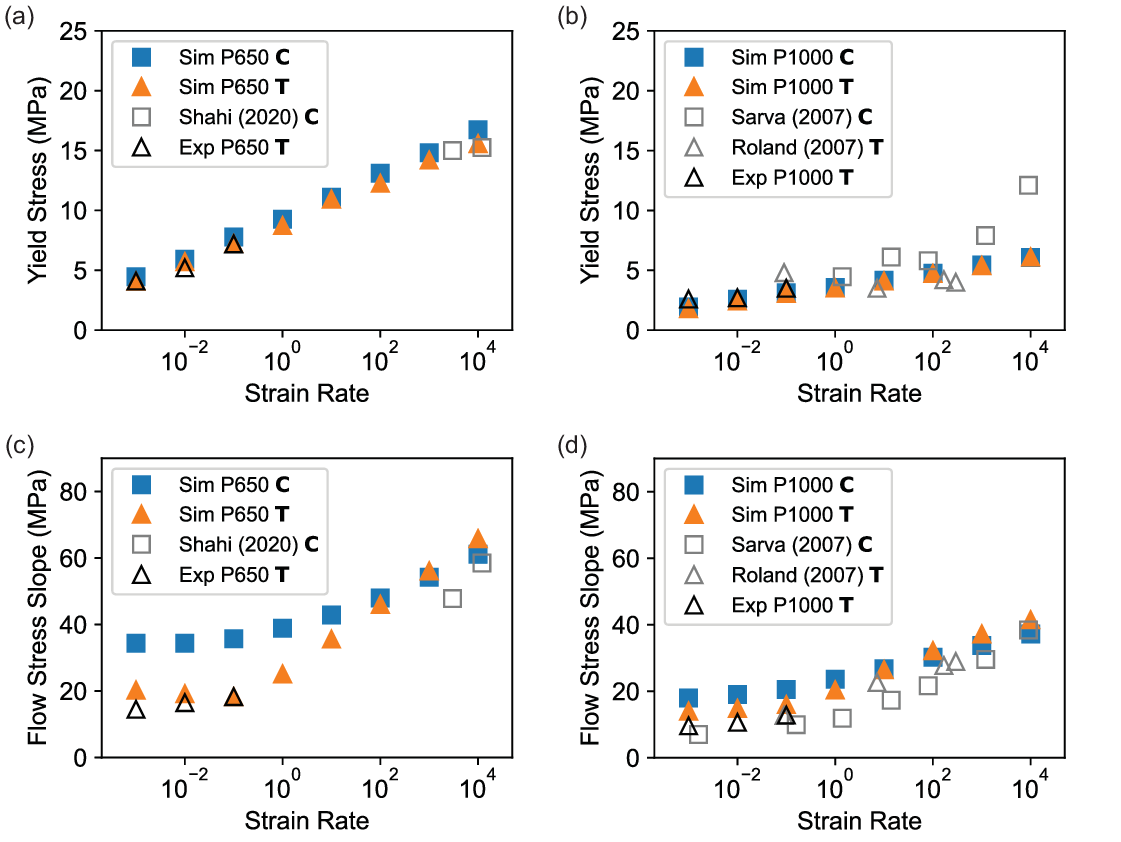}
	\caption{Yield stress vs. strain rate for (a) P650 and (b) P1000. Flow stress slope after yielding vs. strain rate for (c) P650 and (d) P1000. The simulation results are symbolized by blue square markers representing compression loading and orange triangles for tension loading. The experimental data is distinguished by unfilled square markers for compression tests and unfilled triangles for tension tests.}
	\label{FIG:yield_flow_plot}
\end{figure}

To assess the effectiveness of our RVE models in capturing rate dependence, we compare the yield stress and flow stress slope for P650 and P1000 RVE simulations to experimental data from 0.001 to $10^{4}$ $\text{s}^{-1}$ in Figure \ref{FIG:yield_flow_plot}. We select these two quantities because we anticipate that they will exhibit rate dependence, which can be attributed to the rate-dependent viscoplastic flow and hard phase breakdown incorporated in our material model. Literature data does indicate an increase in modulus at high rates \citep{roland_high_2007,shahi_thermo-mechanical_2021,sarva_stressstrain_2007}, but our model is not set up to capture this. Our own data is used for rates between 0.001 and 0.1 $\text{s}^{-1}$. For intermediate and high-rate data, we reference \citet{shahi_thermo-mechanical_2021} for P650, and \citet{sarva_stressstrain_2007} and \citet{roland_high_2007} for P1000. We evaluate our model under both tension and compression, in line with the available experimental data. In both P650 and P1000 RVE simulations, under compression and tension, the yield stress exhibits a linear increase with the logarithm of strain rate. This observation aligns with the viscoplastic flow equation used in our model (Equation \ref{eqn:plastic_flow_2}). We find that the yield stress values for both P650 and P1000 RVEs are similar under tension and compression, with P650 showing a slightly higher yield in compression versus tension. We see that the yield stress of P650 and P1000 is well matched to experimental data at low rates, 0.1 to 0.001 $\text{s}^{-1}$. In addition, we find that the yield stress is well matched for strain rates $10^{3}$ to $10^{4}$ $\text{s}^{-1}$ for P650. For P1000, we see that the yield stress is well matched for intermediate strain rates up to $10^{3}$ $\text{s}^{-1}$. However, for rates past $10^{3}$ $\text{s}^{-1}$, experimental data shows that there is an increased sensitivity to yield stress. To account for the increase in strain rate sensitivity, viscoplastic models of glassy polymers often incorporate two relaxation timescales: the $\beta$-relaxation, which represents short timescales, and the $\alpha$-relaxation, which represents long timescales \citep{mulliken_mechanics_2006}. We opted to represent viscoplastic flow using a single timescale for the sake of simplicity and clarity, which might account for the observed discrepancy in P1000 yield stress.

Examining the flow stress slope after yield, we find that for both P650 and P1000, the RVE simulations and experimental data show that the flow stress slope is relatively constant across low strain rates, followed by an increase at intermediate and high strain rates. In our model, this effect is exclusively attributed to the rate-dependent hard phase breakdown model (Equation \ref{eqn:hard_phase_breakdown_1}), which accurately captures the rate-dependent behavior. There is less hard phase breakdown when the RVE is subjected to higher strain rates (Supplementary Materials Figure S5 for P650 and Figure S6 for P1000). It has been proposed, based on thermomechanical characterization, that the PU undergoes a loading rate-dependent glass transition \citep{shahi_thermo-mechanical_2021,yi_large_2006}. At low rates, PU exhibits a leather-like response, combining features of both elastomers and glassy polymers. However, at high rates, the PU crosses its glass transition temperature, resulting in a more glassy behavior. The observed sensitivity of the flow stress slope could also stem from the soft phase crossing the glass transition during high rates, which our model does not account for. For our model, we see a difference between the flow stress slope under tension and compression, especially at low strain rates for P650. This is due to constraints in our hard phase breakdown evolution (eq. \ref{eqn:hard_phase_breakdown_1}), which require the elastic part of the volumetric change, $J^{e}$, to be greater than 1 for breakdown to occur. Under compression, the RVE lacks the necessary increase in volume to accommodate the density change associated with hard phase breakdown when compared to the RVE under tension loading. Overall, our ability to model key rate-dependent responses using hard phase breakdown and viscoplastic flow highlights the critical role that the hard phase plays in the overall mechanical behavior of polyurea.

To further our understanding of microstructural changes in the hard phase, we performed discrete Fourier transform (FT) analyses of polyurea representative volume elements. Given that we applied the Fourier transform to a spatial domain, the absolute value of the transform provides information about the presence of particular spatial frequencies (domains/nm). Analyzing the spatial frequency of hard domains is particularly significant as it yields information akin to the results from small-angle X-ray scattering (SAXS) experiments \citep{balizer_investigation_2011,castagna_role_2012,castagna_effect_2013,choi_microstructure_2012,pangon_influence_2014,rinaldi_microstructure_2011,rosenbloom_microstructural_2021}. For these analyses, we employed RVE simulations with 60 nm in each dimension, instead of the previously discussed RVEs that are 10 nm in each dimension. The larger size RVE enhances the resolution of our FT analysis. The stress-strain responses of the small and large RVE sizes closely match, which confirms that the 10 nm RVE adequately represents the microstructure of PU (Supplementary Materials Figure S7). For clarity, we will focus our discussion on the P650 RVE. The P1000 RVE results are shown in the supplementary material Figure S8. Figure \ref{FIG:P650_RVE_FT_SAXS}a shows the P650 RVE deformed to various strain levels. When the RVE is undeformed and at low strains ($\sim$0.1), the hard phase is relatively continuous. However, as strain increases to around 0.3, the hard phase fragments into smaller pieces. By the time strain reaches 0.5, the hard phase is primarily dispersed into smaller domains. Significantly, the hard phase breakup captured by our finite element RVE and constitutive model agrees with a coarse-grained non-equilibrium MD study performed by \citet{zheng_molecular_2022}. The authors found that the hard phase was continuous at strains less than 0.2 and then broke up into smaller pieces after 0.4 strain.

\begin{figure}[htb!]
	\centering
    \includegraphics[scale=0.8]{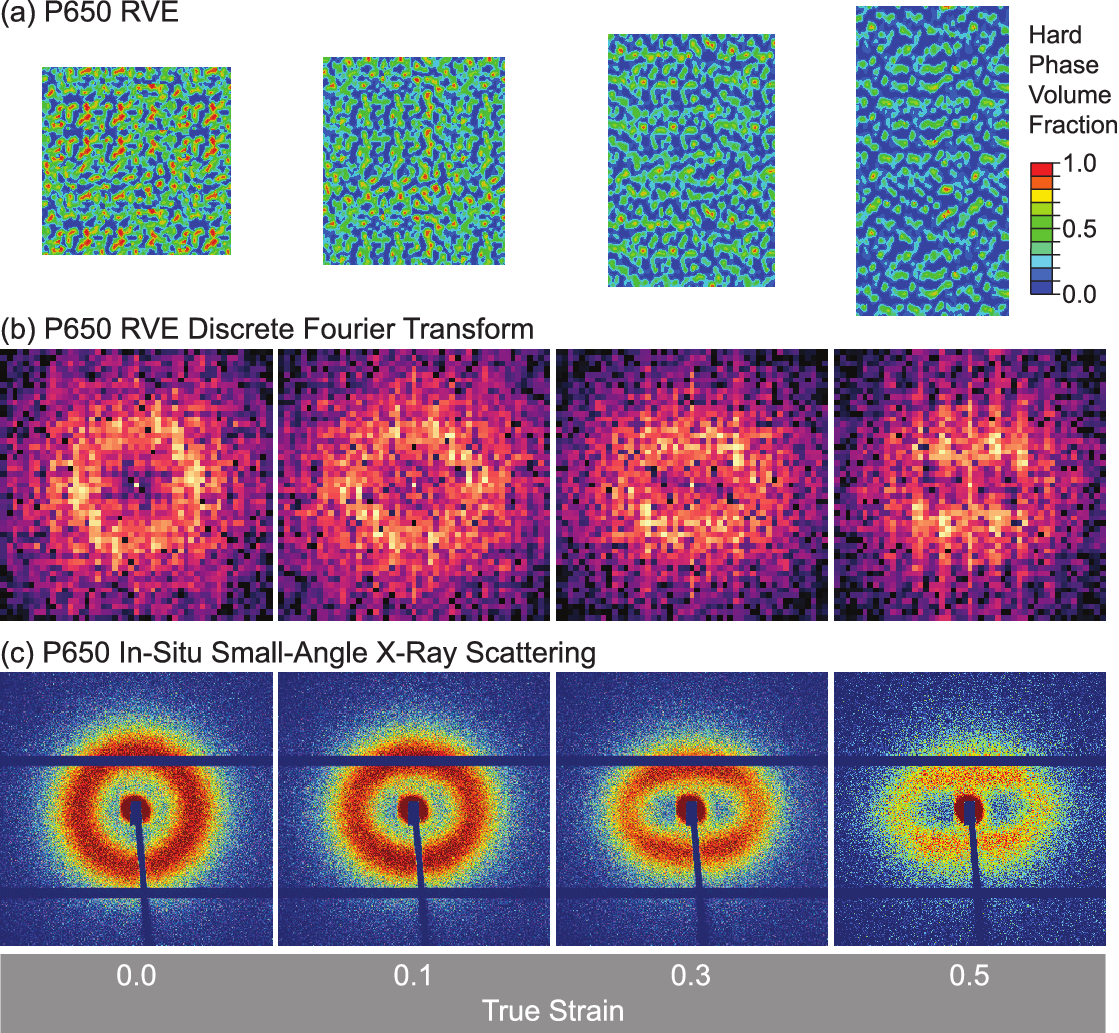}
	\caption{(a) Snapshots of the 60x60x60 $\text{nm}^{\text{3}}$ P650 RVE deformed to various true strain levels. (b) Centrally positioned 2D cross-sections within the 3D Fourier transformation of the P650 RVE. (c) In situ small angle X-ray scattering of P650 during tensile deformation.}
	\label{FIG:P650_RVE_FT_SAXS}
\end{figure}

Figure \ref{FIG:P650_RVE_FT_SAXS}b shows the 2D cross sections in the middle of the Fourier space at various strain levels. The vertical axis (meridian) aligns with the loading direction, while the horizontal axis (equator) aligns with the transverse direction. Regular spacing of the amorphous hard phase causes a broad, high-amplitude region in the transform. At zero strain, the domain spacing appears relatively isotropic, evident as a circular ring in the FT plots. As the RVE undergoes deformation, the circular ring transforms into an ellipse, with its minor axis aligning with the loading direction. This indicates an increase in domain spacing along the loading direction and a decrease in the transverse direction. By 0.5 strain, there is a relative increase in meridional intensity and a decrease in equatorial intensity into a two-point pattern. In Figure \ref{FIG:P650_RVE_FT_SAXS}c, we show experimentally obtained 2D scattering patterns for P650 from SAXS. It is important to note that the scale of the FT and SAXS images differ. Nonetheless, these SAXS patterns reflect our simulation results. The deformation process sees a similar evolution from a circular to an elliptical pattern. In addition, for P650, there is a rise in meridional intensity at 0.5 strain, consistent with our simulated two-point pattern. 

The observed two-point pattern in SAXS experiments is suggested to result from the elongation of the hard domains \citep{choi_microstructure_2012,rinaldi_microstructure_2011}. Specifically, a meridional two-point pattern is associated with horizontally arranged layers \citep{murthy_deformation_2002,murthy_small-angle_2021}. Our RVE simulation and Fourier transform analysis further support this idea. When examining the RVE snapshot at 0.5 strain, it becomes apparent that many hard domains are horizontally elongated while fewer are oriented at an angle. There are minimal elongated domains in the vertical direction. In our simulations, this can be attributed to a greater reduction of the hard phase between hard domains along the loading direction and a lesser reduction in the direction perpendicular to the load. It is worth noting that elongated phases oriented at an angle cause a four-point scattering pattern where the points are located in each quadrant of the plane \citep{murthy_deformation_2002,murthy_small-angle_2021}. We observe a faint four-point pattern in our P1000 SAXS image at 0.3 strain (Supplementary Materials S8). This is in accordance with the four-point pattern reported by \citet{rinaldi_microstructure_2011} for PU P1000. While the FT of our P1000 RVE potentially has higher intensities in the four-point pattern, it lacks sufficient resolution for a definite conclusion. Nonetheless, an inspection of the simulated P1000 RVE at 0.5 strain shows primarily both non-elongated and diagonally oriented domains.

\begin{figure}[htb!]
	\centering
    \includegraphics[scale=0.75]{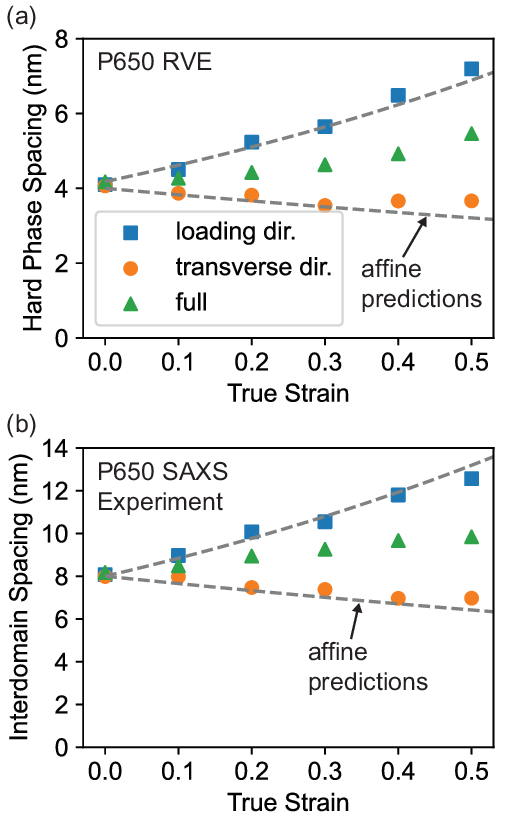}
	\caption{(a) Hard phase spacing of P650 RVE simulations (b) Bragg interdomain spacing of P650 from small-angle X-ray scattering experiments. The dashed gray lines represent the affine prediction of domain spacing in the loading and transverse directions.}
	\label{FIG:P650_domain_spacing}
\end{figure}

To more effectively analyze the domain deformation, we measure the peak location in the loading direction, transverse directions, and overall azimuthal average of the RVE Fourier transform and SAXS data. From the peak location, we can calculate the spacing between the hard phases of the RVE (Figure \ref{FIG:P650_domain_spacing}a) and the Bragg interdomain spacing from the SAXS data (Figure \ref{FIG:P650_domain_spacing}b). The simulated P650 hard phase spacing starts as isotropic, with spacing across all directions at $\sim$4.1 nm. This closely matches our CGMD hard bead radial distribution of $\sim$4.2 nm. The SAXS measurement of PU also indicates as isotropic albeit with a larger interdomain spacing of $\sim$8 nm. The difference between our simulated and experimentally measured spacing may result from discrepancies between our CGMD model and the actual polymer, as discussed in Section \ref{sec:results_md}. Moreover, our SAXS domain spacing of $\sim$8 nm for P650 and $\sim$9 nm for P1000 (Supplementary Materials Figure S9) is somewhat higher than previously reported values for P650 being 6-7 nm and P1000 being 7-8 nm \citep{pangon_influence_2014,castagna_effect_2013,rinaldi_microstructure_2011,choi_microstructure_2012}. Our larger domain spacing could be due to solvent use during casting or heating the solidified polymers for 48 hrs to remove traces of solvent.

Figure \ref{FIG:P650_domain_spacing} illustrates that deformation leads to increased spacing in the loading direction and decreased spacing in the transverse direction according to both the RVE hard phase spacing and SAXS measurements. The slight spacing increase in the full azimuthal average is a consequence of these combined effects. The spacing follows the affine deformation predictions for both the simulation and experiment. The transverse prediction uses a Poisson's ratio of 0.44, obtained from the simulated RVE. Notably, this simulated Poisson's ratio effectively predicts the reduction in transverse domain spacing seen in the SAXS experiments. Critically, the domain spacing remains affine despite strain localization and hard phase breakdown in the RVE simulation. This indicates that most of the strain localization and breakdown happens at the domain boundaries, while the centers of the domains experience less change. Overall, the alignment of our Fourier transform analysis with SAXS experiments underscores the effectiveness of our RVE method and constitutive model in capturing key microstructural changes in polyurea.

\section{Conclusion}
\label{sec:conclusion}

This work centers on the significant impact of the phase-segregated microstructure of polyurea on its mechanical properties. We used coarse-grained molecular dynamics (CGMD) simulations to create representative volume elements (RVE) for two polyurea compositions with different soft segment lengths: P650 and P1000. In the RVE, each material point was assigned a certain volume fraction of the hard and soft phases. Following this, we utilized these RVEs in a finite element simulation framework, which enabled us to model polyurea's behavior under various loading conditions. In our unique approach, we used separate constitutive models for the soft and hard phases where the volume fractions in the RVE determined the relative contribution of the phases at each material point. The soft phase was modeled as an elastomer, while the hard phase was modeled as an elastic-viscoplastic glassy polymer. Notably, we incorporated the concept of deformation-induced breakdown in the hard phase. We modeled a plasticity-driven mechanism of hard phase break down into a separate phase with properties resembling the soft phase. Importantly, we modeled a density reduction during the breakdown, hypothesizing that the densely packed glassy hard phase intermixes with the elastomeric soft phase, causing this decrease. The density decrease required us to create restrictions on what local deformation conditions can lead to breakdown in order to prevent undue strain energy increases during the process. 

The only parameters that we varied between our P650 and P1000 RVE simulations were the RVE microstructure obtained from CGMD and soft phase material properties estimated from the polymer's chain density and length obtained from experimental gel permeation chromatography measurements. All other material properties, such as the hard phase modulus, viscoplastic flow, and breakdown properties, were calibrated to match P650 experimental data. Remarkably, the differences in RVE microstructure and soft phase properties allow us to predict key features from P1000 experimental results. In addition, our simulations show the significance of polyurea's heterogeneous microstructure in shaping the stress-strain profiles during cyclic loading. This is evident when comparing our RVE simulations to homogeneous simulations with equivalent average hard phase volume fractions. Moreover, our RVE models can predict the yield stress and post-yield flow stress slopes across a wide range of strain rates for both P650 and P1000. One consequence of incorporating density changes during hard phase breakdown and accounting for associated deformation requirements for breakdown to occur is the simulated difference in flow stress between tension and compression loading. It remains to be seen whether the actual polymer would exhibit this effect since there is a need for a comprehensive data set with both P650 and P1000 subjected to compression and tension loading across the wide range of strain rates we simulated. Our model successfully captures the rate-dependent effects with the viscoplastic flow of the hard phase and rate-dependent hard phase breakdown. The experimentally observed rise in flow stress and yield at high strain rates might be attributed to a rate-dependent glass transition of the soft phase, which we have not modeled. Further work could involve modeling the rate-dependent effects of the soft phase. 

Our constitutive model and RVE simulations capture key microstructural behaviors of polyurea. By incorporating hard phase breakdown in our simulations, we observe the transformation of initially continuous hard domains into dispersed fragments within our simulations. Moreover, conducting a Fourier transform analysis on the RVE allows us to compare our results with small-angle X-ray scattering experiments. Both the RVE and SAXS exhibit affine deformation in terms of hard domain spacing, confirming the occurrence of hard phase breakdown at domain interfaces. The Fourier transform of our P650 simulation displays a two-point pattern observed in SAXS measurements, suggesting that the hard phase undergoes fibrillation — a behavior accurately captured by our hard phase breakdown model. The discrepancy between the domain spacing observed in our simulations and the measurements obtained from SAXS arises from the initial simulated microstructure derived from coarse-grained molecular dynamics. Consequently, further investigation is warranted to develop molecular dynamics models that accurately reproduce the interdomain spacing observed in experimental measurements. Lastly, although our plasticity-driven hard phase breakdown model successfully captured key behaviors observed in experiments, we believe further work can be done to develop nanophase breakdown models in polymers within a finite element or finite volume simulation framework. Specifically, models that consider non-local information, such as the quantities of phases surrounding each material point. In conclusion, we anticipate that the modeling and analysis framework presented will serve as a method for understanding the mechanical and microstructural behavior of phase-segregated polymers. Applying this method to other phase-segregated polymers would require steps such as creating a microstructure through simulation or imaging, converting this microstructure into volume fractions on an RVE grid, and finally determining the suitable constitutive models for each phase.


\section{Acknowledgments}

This work was supported by the Office of Naval Research under Grant N00014-19-1-2099 administered by PO Dr. Barsoum. Allocation TG-MSS140006 on the Texas Advanced Computing Center Stampede2 cluster was used for all simulations. We thank Bayville Chemical Supply Company Inc. for providing the Dow Isonate 143L and Evonik Industries AG for providing the Versalink. Small-angle X-ray scattering experiments were conducted at the Materials Solutions Network at Cornell High Energy Synchrotron Source (MSN-C), which is supported by the Air Force Research Laboratory under award FA8650-19-2-5220. We thank Louisa Smieska, Zhongwu Wang, and Arthur Woll for their assistance at the Cornell High Energy Synchrotron Source. 

\section{Declaration of Generative AI and AI-assisted technologies in the writing process}

During the preparation of this work the authors used ChatGPT 3.5 and 4.0 (OpenAI) in order to perform language editing. In addition the authors used Grammarly (Grammarly Inc.) for grammar and spell checking. After using these tools and services, the authors reviewed and edited the content as needed and take full responsibility for the content of the publication.

\bibliographystyle{elsarticle-harv} 

\pagebreak
\begin{center}
\textbf{\Large Supplemental Materials: Elucidating the impact of microstructure on mechanical properties of phase-segregated polyurea: Finite element modeling of molecular dynamics derived microstructures}
\end{center}

\setcounter{equation}{0}
\setcounter{section}{0}
\setcounter{figure}{0}
\setcounter{table}{0}
\makeatletter
\renewcommand{\theequation}{S\arabic{equation}}
\renewcommand{\thefigure}{S\arabic{figure}}
\renewcommand{\thesection}{S\arabic{section}}
\renewcommand{\thetable}{S\arabic{table}}
\renewcommand{\bibnumfmt}[1]{[S#1]}
\renewcommand{\citenumfont}[1]{S#1}

\noindent{\bfseries{\large Contents}\hfill \par}
\contentsline {section}{\numberline {S1}Determining soft segment molecular weight for Versalink P650 and P1000}{\pageref{sec:si_soft_segment_mw}}{section}
\contentsline {section}{\numberline {S2}Hard phase breakdown dissipation inequality addendum}{\pageref{sec:si_hard_breakdown}}{section}
\contentsline {section}{\numberline {S3}Finite element RVE 10x10x10 nm$^{\text{3}}$}{\pageref{sec:si_FE_RVE}}{section}
\contentsline {section}{\numberline {S4}Hard phase breakdown for homogeneous simulations}{\pageref{sec:si_hom_hard_phase_breakdown}}{section}
\contentsline {section}{\numberline {S5}P650 monotonic loading}{\pageref{sec:si_P650_monotonic}}{section}
\contentsline {section}{\numberline {S6}P1000 monotonic loading}{\pageref{sec:si_P1000_monotonic}}{section}
\contentsline {section}{\numberline {S7}Stress-strain for 10x10x10 nm$^{\text{3}}$ RVE vs. 60x60x60 nm$^{\text{3}}$ RVE}{\pageref{sec:si_RVE_size_comparison}}{section}
\contentsline {section}{\numberline {S8}P1000 60x60x60 nm$^{\text{3}}$ RVE rendering, RVE Fourier transform, SAXS}{\pageref{sec:si_P1000_FT_saxs}}{section}
\contentsline {section}{\numberline {S9}P1000 domain spacing}{\pageref{sec:si_P1000_domain_spacing}}{section}

\section{Determining soft segment molecular weight for Versalink P650 and P1000}
\label{sec:si_soft_segment_mw}
\begin{figure}[htb!]
	\centering
    \includegraphics[scale=0.8]{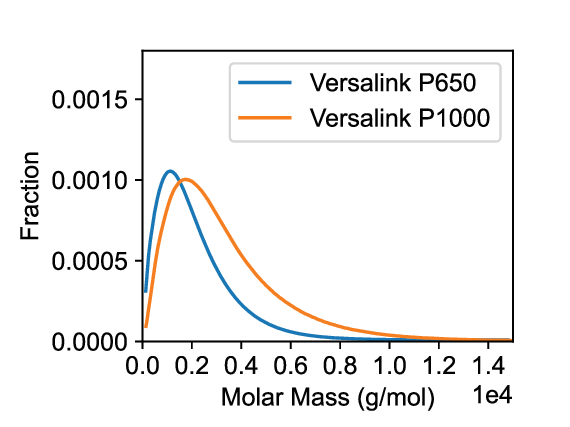}
	\caption{Molecular weight distribution for Versalink P650 and P1000.}
	    \label{fig:gpc_distribution}
\end{figure}

We employed gel permeation chromatography (GPC) using a Tosoh EcoSec 9320GPC system with SuperHM-M columns with a series flow rate of 0.350 mL/min to measure the molecular weight of Versalink P650 and P1000 oligomers. THF was used as the solvent and eluent. The number average ($\text{M}_{\text{n}}$), weight average ($\text{M}_{\text{w}}$), and z average ($\text{M}_{\text{z}}$) molecular weights were determined by light scattering with a Wyatt miniDawn Treos multi-angle light scattering detector. The calibration curve for molecular weight was established using TSKgel polystyrene standards. The molecular weight distributions are shown in Figure \ref{fig:gpc_distribution} and Table \ref{tbl:versalink_molecular_weight} shows the obtained values. 

\begin{table}[htb!]
\centering
\caption{A table of molecular weights for Versalink P650 and P1000. $M_{n}$ is the number average, $M_{w}$ is the weight average, and $M_{z}$ is the z average molecular weight.}
\label{tbl:versalink_molecular_weight}
\begin{tabular}{llll}
\hline
& $M_{n}$ (g/mol) & $M_{w}$ (g/mol) & $M_{z}$ (g/mol)   \\\hline
P650 & 573 & 1211 & 2334 \\\hline
P1000 & 920 & 1918 & 3374 \\\hline
\end{tabular}
\end{table}

\begin{figure}[H]
	\centering
    \includegraphics[scale=1.0]{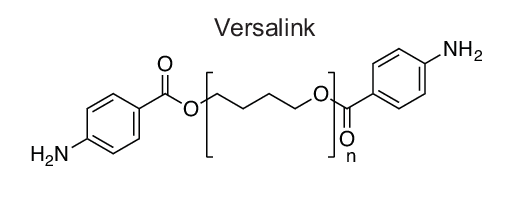}
	\caption{Chemical structure of Versalink}    \label{fig:versalink_structure}
\end{figure}

The Versalink oligomers contain aromatic end groups as shown in Figure \ref{fig:versalink_structure}. 
In our coarse-grained MD simulations, the soft beads correspond to the poly(tetrahydrofuran) repeat units. To find the molecular weight of the soft segments, we subtracted the combined molecular weight of the end groups (258.28 g/mol). By subtracting the end group weight, we obtained the weight-averaged molecular weights of P650 and P1000, which are 952.7 g/mol and 1659.7 g/mol, respectively. 

\section{Hard phase breakdown dissipation inequality addendum}
\label{sec:si_hard_breakdown}

The derivation below shows that
\begin{equation}
\label{eqn:dissipation_ineq4_2}
\frac{\partial}{\partial \xi_{h0}}\Big[  \phi_{h}\xi_{h0}(W_{h0^{\alpha}}^{vol} + W_{h0^{\beta}}^{vol}) + \phi_{h}(1 - \xi_{h0})(W_{h1}^{vol})  + (1 - \phi_{h})(W_{s}^{vol})  \Big] \geq 0
\end{equation}
is satisfied when $J^{e} > 1$, provided that $(k_{h0^{\alpha}} + k_{h0^{\beta}}) \geq k_{h1}$ and $\rho_{h0} \geq \rho_{h1}$. Satisfying Inequality \ref{eqn:dissipation_ineq4_2} ensures no undue strain energy density increase during the breakdown process. We begin by substituting the volumetric strain energy functions for mechanisms $h0^{\alpha}$, $h0^{\beta}$, $h1$, and $s$ (as described in the main article) into inequality \ref{eqn:dissipation_ineq4}. It is worth noting that $J^{e}$ corresponds to $J/{J}^{h0 \rightarrow h1}$.
\begin{equation}
\frac{\partial}{\partial \xi_{h0}}  \Bigg[ \frac{1}{2} \phi_{h} \Big(\xi_{h0}k_{h0^{\alpha}} + \xi_{h0}k_{h0^{\beta}}  + k_{h1} - \xi_{h0}k_{h1} \Big) \Big(\frac{J}{{J}^{h0 \rightarrow h1}} - 1 \Big)^{2} +  \frac{1}{2}\big(1 - \phi_{h} \big)k_{s} \big(J - 1 \big)^{2} \Bigg] \geq 0.
\end{equation}
Noting that ${J}^{h0 \rightarrow h1}$ is a function of $\xi_{h0}$, we use the derivative product rule and chain rule to obtain the following inequality: 
\begin{multline}
\label{eqn:A2}
\frac{1}{2}\Big[ \Big(k_{h0^{\alpha}} + k_{h0^{\beta}} - k_{h1} \Big) \Big(\frac{J}{J^{h0 \rightarrow h1}} - 1 \Big)^{2}  \\
+ \Big(\xi_{h0}k_{h0^{\alpha}} + \xi_{h0}k_{h0^{\beta}}  + k_{h1} - \xi_{h0}k_{h1} \Big) \frac{\partial}{\partial J^{h0 \rightarrow h1}}\Big(\frac{J}{J^{h0 \rightarrow h1}} - 1  \Big)^{2}    \frac{\partial}{\partial \xi_{h0}}  J^{h0 \rightarrow h1}    \Big] \geq 0.
\end{multline}
By substituting the function for ${J}^{h0 \rightarrow h1}$ (as described in the main article) into inequality \ref{eqn:A2} and evaluating the derivatives, we obtain the following inequality:
\begin{multline}
\label{eqn:A3}
\frac{1}{2}\Bigg[ \Big(k_{h0^{\alpha}} + k_{h0^{\beta}} - k_{h1} \Big) \Big(\frac{J}{J^{h0 \rightarrow h1}} - 1 \Big)^{2}  \\
+ \Big(\xi_{h0}k_{h0^{\alpha}} + \xi_{h0}k_{h0^{\beta}}  + k_{h1} - \xi_{h0}k_{h1} \Big) \Big( \frac{2J(J^{h0 \rightarrow h1} - J)}{J^3}\Big)  \Bigg(   \frac{\rho_{h0}(\rho_{h1} - \rho_{h0})}{\big(\rho_{h0}\xi_{h0} + \rho_{h1}(1-\xi_{h0})     \big)^{2}}  \Bigg)  \Bigg] \geq 0.
\end{multline}
If we specify that $(k_{h0^{\alpha}} + k_{h0^{\beta}}) \geq k_{h1}$ the first term in equation \ref{eqn:A3} is positive. In addition, if we specify that $\rho_{h0} \geq \rho_{h1}$, the second term is positive if $J \geq J^{h0 \rightarrow h1}$. This is equivalent to $J^{e} \geq 1$. 

\newpage

\section{Finite element RVE 10x10x10 nm$^3$}
\label{sec:si_FE_RVE}
\begin{figure}[htb!]
	\centering
    \includegraphics[scale=1.0]{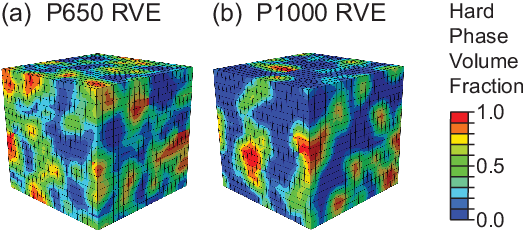}
	\caption{Snapshots of (a) P650 and (b) P1000 representative volume elements for 10x10x10 nm$^3$ finite element simulations}   \label{fig:rve_snaps}
\end{figure}

\section{Hard phase breakdown for homogeneous simulations}
\label{sec:si_hom_hard_phase_breakdown}

\begin{figure}[htb!]
	\centering
    \includegraphics[scale=0.8]{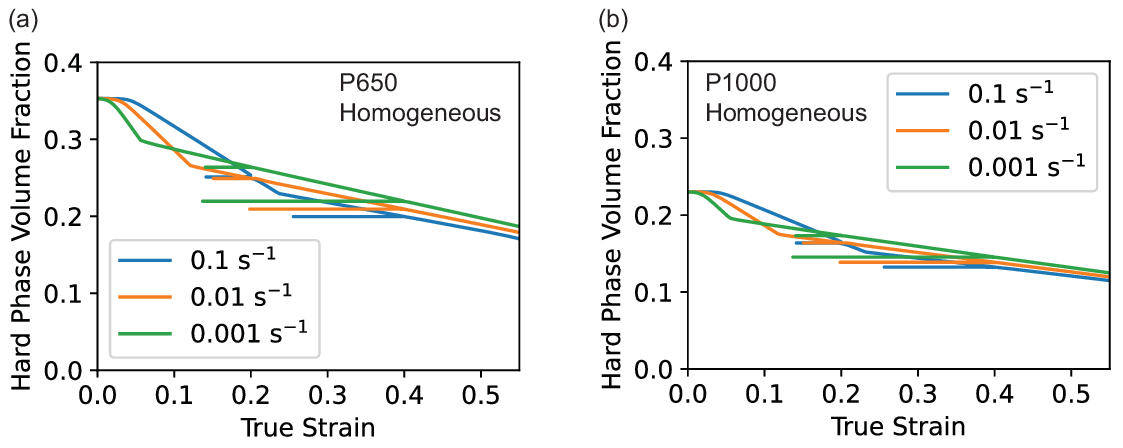}
	\caption{Hard phase volume fraction vs. strain for (a) P650 and (b) P1000 homogeneous simulations undergoing cyclic loading.}   \label{fig:hom_hard_phase_breakdown}
\end{figure}

\newpage
\section{P650 monotonic loading}
\label{sec:si_P650_monotonic}

\begin{figure}[htb!]
	\centering
    \includegraphics[scale=0.8]{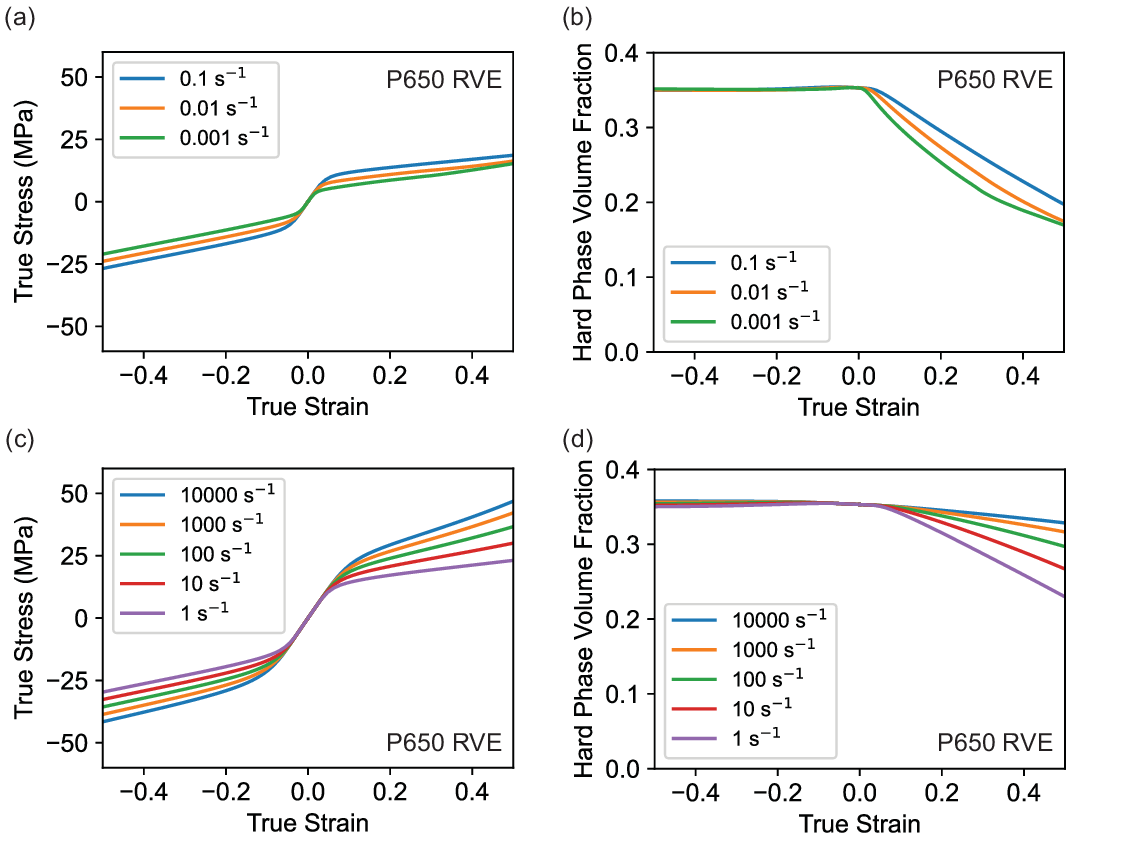}
	\caption{Monotonic loading simulations of the P650 RVE. (a) True stress vs. true strain, and (b) hard phase volume fraction vs. true strain for low-rate loading. (c) True stress vs. true strain, and (b) hard phase volume fraction vs. true strain for intermediate and high-rate loading.}   
    \label{fig:P650_monotonic_stress_vf}
\end{figure}

\newpage
\section{P1000 monotonic loading}
\label{sec:si_P1000_monotonic}

\begin{figure}[htb!]
	\centering
    \includegraphics[scale=0.8]{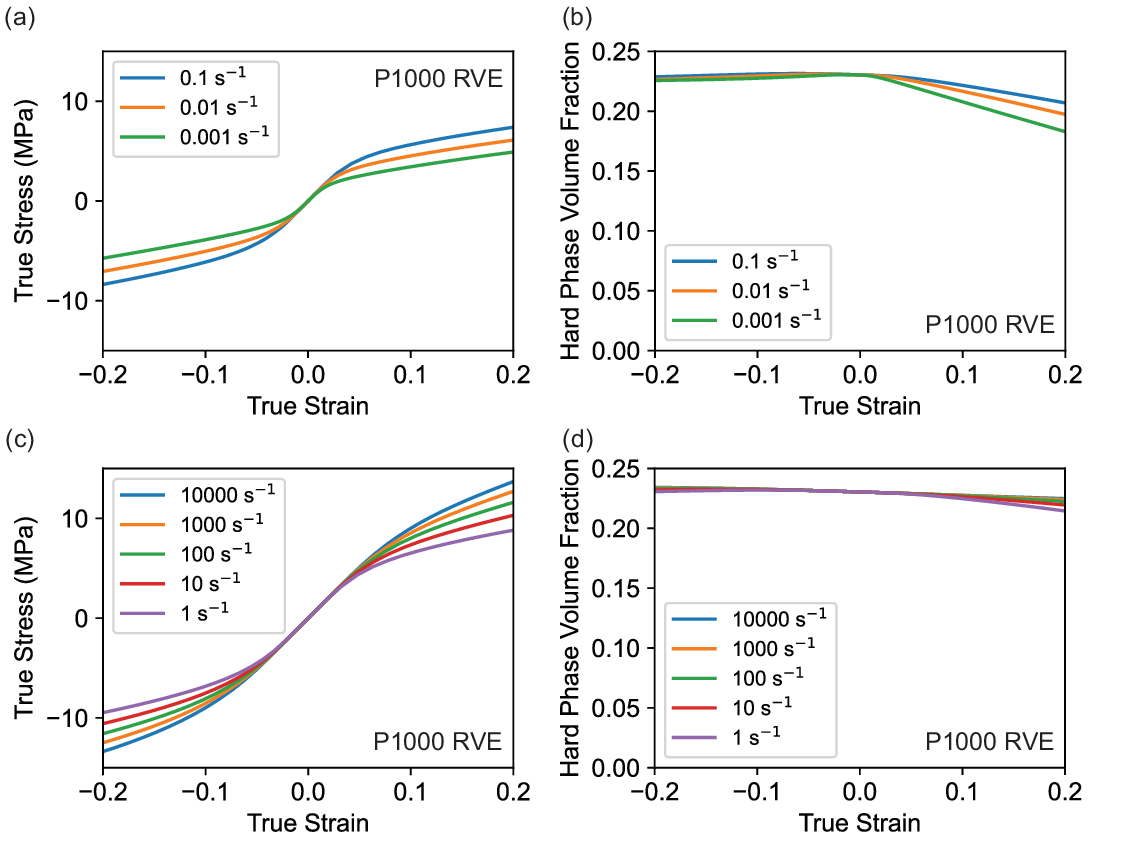}
	\caption{Monotonic loading simulations of the P1000 RVE. (a) True stress vs. true strain, and (b) hard phase volume fraction vs. true strain for low-rate loading. (c) True stress vs. true strain, and (b) hard phase volume fraction vs. true strain for intermediate and high-rate loading.}   
    \label{fig:P1000_monotonic_stress_vf}
\end{figure}

\newpage

\section{Stress-strain for 10x10x10 nm$^{\text{3}}$ RVE vs. 60x60x60 nm$^{\text{3}}$ RVE}
\label{sec:si_RVE_size_comparison}

\begin{figure}[htb!]
	\centering
    \includegraphics[scale=0.8]{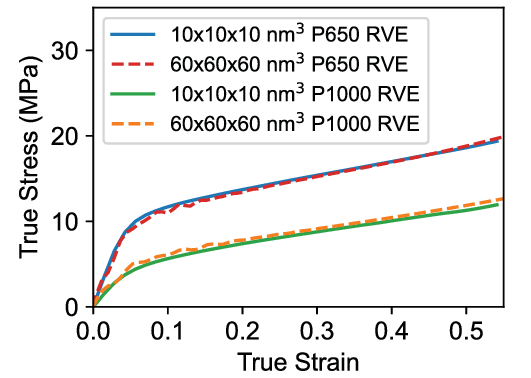}
	\caption{Comparison of true stress vs. true strain curves for the 10x10x10 $\text{nm}^{\text{3}}$ RVE and 60x60x60 $\text{nm}^{\text{3}}$ RVE. Both P650 and P1000 RVE simulations are shown in this figure.}   
    \label{fig:rve_size_comparison}
\end{figure}

\newpage

\section{P1000 60x60x60 nm$^{\text{3}}$ RVE rendering, RVE Fourier transform, SAXS}
\label{sec:si_P1000_FT_saxs}

\begin{figure}[htb!]
	\centering
    \includegraphics[scale=0.8]{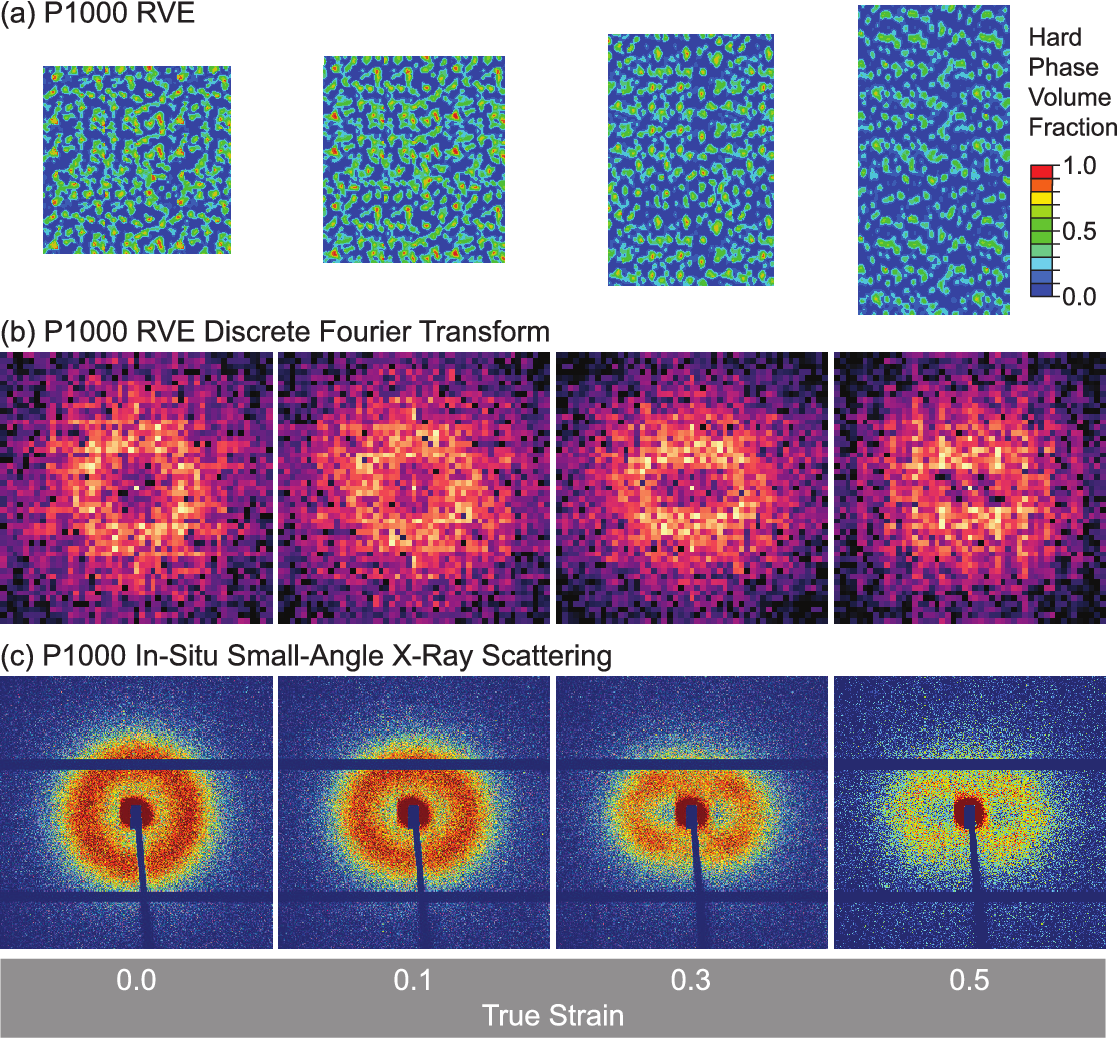}
	\caption{(a) Snapshots of the 60x60x60 $\text{nm}^{\text{3}}$ P1000 RVE deformed to various true strain levels. (b) Centrally positioned 2D cross sections within the 3D Fourier transformation of the P1000 RVE. (c) In-situ small angle X-ray scattering of P1000 during tensile deformation.}   
    \label{fig:P1000_RVE_FT_SAXS}
\end{figure}

\newpage

\section{P1000 domain spacing}
\label{sec:si_P1000_domain_spacing}

\begin{figure}[htb!]
	\centering
    \includegraphics[scale=0.8]{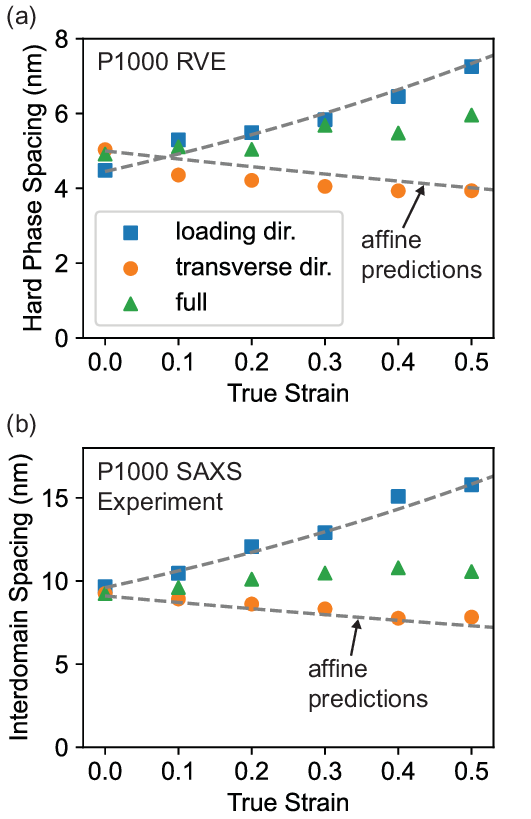}
	\caption{(a) Hard phase spacing of P1000 RVE simulations (b) Bragg interdomain spacing of P1000 from small-angle X-ray scattering experiments. The dashed gray lines represent the affine prediction of domain spacing in the loading and transverse directions.}
    \label{fig:P1000_RVE_Domain_Spacing}
\end{figure}





\end{document}